\documentclass[10.5pt, twoside]{article} 

\usepackage{natbib}
\usepackage{array, epsfig, rotating}
\usepackage{graphicx}
\usepackage[top=1in, bottom=1in, left=1in, right=1in]{geometry}
\usepackage{hyperref}

\usepackage[nodisplayskipstretch]{setspace}

\newcommand{\bib}{\par\noindent\hangindent=0.5 true cm\hangafter=1}

\usepackage{amsmath, amssymb, amsfonts,amsthm}
\usepackage{multirow}
\usepackage{fancyhdr}
\usepackage{setspace}
\usepackage{etoolbox}
\usepackage{caption, subcaption}
\usepackage{titlesec}
\patchcmd{\thebibliography}{\chapter*}{\section*}{}{}
\makeatletter
\patchcmd{\chapter}{\if@openright\cleardoublepage\else\clearpage\fi}{}{}{}
\makeatother

\titlespacing*{\section}
{0pt}{3ex plus .5ex minus .2ex}{1ex plus .2ex}
\titlespacing*{\subsection}
{0pt}{3ex plus .5ex minus .2ex}{1ex plus .2ex}
\newtheorem*{app-thm}{Theorem}

\newenvironment{proof.}{\noindent{\bf Proof.}\hspace*{1ex}}{\qed\\}

\hypersetup{
    colorlinks=true,        
    linkcolor=red,          
    citecolor=blue,         
    urlcolor=violet         
}

\def\mP_n{\mathbb{P}_n}

\renewcommand\thesection{\arabic{section}}

\def\beq{\begin{eqnarray}}
\def\eeq{\end{eqnarray}}

\def\beqn{\begin{eqnarray*}}  
\def\eeqn{\end{eqnarray*}}

\def\saveusalittle{{\vspace{-0.3cm}}}
\def\saveustwolittle{{\vspace{-0.5cm}}}

\def\E{{\rm E}}
\def\Var{{\rm Var}}

\def\dd{{\rm d}}
\def\N{{\rm N}}

\def\Pr{P}
\def\pr{{\rm pr}}

\def\Beta{{\rm Beta}}
\def\Gamm{{\rm Gamma}}

\def\KL{{\rm KL}}

\def\quadandquad{\quad {\rm and} \quad}
\def\arr{\rightarrow}
\def\hatt{\widehat}
\def\tilda{\widetilde}
\def\sumin{\sum_{i=1}^n}
\def\prodin{\prod_{i=1}^n}
\def\maxin{\max_{i\le n}}
\def\eps{\varepsilon}
\def\half{\hbox{$1\over2$}}
\def\third{\hbox{$1\over3$}}

\def\rootn{\sqrt{n}}

\def\tr{{\rm t}}
\def\dell{\partial}

\def\prof{{\rm prof}}
\def\KL{{\rm KL}}
\def\fic{{\rm fic}}

\def\argmax{{\rm argmax}}
\def\true{{\rm true}}

\def\mse{{\rm mse}}

\def\cc{{\rm cc}}

\def\pl{{\rm ml}} 
\def\hl{{\rm hl}}
\def\el{{\rm el}} 
\def\narr{{\rm narr}}
\def\wide{{\rm wide}}
\def\dellmudelltheta{\frac{\dell\mu}{\dell\theta}}
\def\dellone{\hbox{$\dell\psi\over \dell\theta$}}
\def\delltwo{\hbox{$\dell\psi\over \dell\gamma$}}
\def\plus{{\rm plus}}

\date{21-June-2017}

\theoremstyle{theorem}
\newtheorem{theorem}{Theorem}
\newtheorem{corollary}{Corollary}
\newtheorem{lemma}{Lemma}

\newtheorem{remark}{Remark}
\newtheorem{example}{Example}

\begin{document}




\renewenvironment{abstract}
 {\small
  \list{}{
    \setlength{\leftmargin}{0.5in}%
    \setlength{\rightmargin}{\leftmargin}%
  }%
  \item\relax}
 {\endlist}

\thispagestyle{empty}
\fontsize{12}{14pt plus.8pt minus .6pt}\selectfont
\vspace{-0.3in}
\centerline{\Large\bf Hybrid combinations of parametric 
   and empirical likelihoods} \vspace{.5cm}
\centerline{Nils Lid Hjort\footnote{N.L.~Hjort is supported 
via the Norwegian Research Council funded project FocuStat.}, 
Ian W. McKeague\footnote{I.W. McKeague is partially supported 
by NIH Grant R01GM095722. }, 
and Ingrid Van Keilegom\footnote{I. Van Keilegom is financially 
supported by the European Research Council (2016-2021, Horizon 2020, 
grant agreement No.\ 694409), and by IAP research network 
grant nr.\ P7/06 of the Belgian government (Belgian Science Policy). }} \vspace{.3cm}
\centerline{\it University of Oslo, Columbia University, 
and KU Leuven} \vspace{.3cm} \fontsize{9}{11.5pt
plus.8pt minus .6pt}\selectfont
\setlength{\parskip}{0.3\baselineskip}

\centerline{July 2017} 

\begin{abstract}
\onehalfspacing \noindent {\it Abstract:} 
This paper develops a hybrid likelihood (HL) method based 
on a compromise between parametric and nonparametric 
likelihoods. Consider the setting of a parametric model 
for the distribution of an observation $Y$ with parameter 
$\theta$. Suppose there is also an estimating function 
$m(\cdot,\mu)$ identifying another parameter $\mu$ via 
$\E\,m(Y,\mu)=0$, at the outset defined independently 
of the parametric model. To borrow strength from the 
parametric model while obtaining a degree of robustness from 
the empirical likelihood method, we formulate inference 
about $\theta$ in terms of the hybrid likelihood function 
$H_n(\theta)=L_n(\theta)^{1-a}R_n(\mu(\theta))^a$. 
Here $a\in[0,1)$ represents the extent of the compromise, 
$L_n$ is the ordinary parametric likelihood for $\theta$, 
$R_n$ is the empirical likelihood function, and $\mu$  
is considered through the lens of the parametric model. 
We establish asymptotic normality of the corresponding HL 
estimator and a version of the Wilks theorem. 
We also examine extensions of these results under 
misspecification of the parametric model, and propose 
methods for selecting the balance parameter $a$.  

\vspace{5pt}
\par
\end{abstract}
\par


\fontsize{12}{14pt plus.8pt minus .6pt}\selectfont

\setcounter{section}{0} 
\setcounter{equation}{0} 
\setcounter{page}{1}
\doublespacing


\baselineskip20pt 
\parskip0pt


\section*{Some personal reflections on Peter}

We are all grateful to Peter for his deeply influential 
contributions to the field of statistics, in particular 
to the areas of nonparametric smoothing, bootstrap, empirical likelihood 
(what this paper is about), functional data, high-dimensional data, 
measurement errors, etc., many of which were major breakthroughs 
in the area. His services to the profession were also exemplary 
and exceptional. It seems that he could simply not say `no' 
to the many requests for recommendation letters, thesis reports, 
editorial duties, departmental reviews, and various other 
requests for help, and as many of us have experienced, he handled 
all this with an amazing speed, thoroughness, and efficiency. 
We will also remember Peter as a very warm, gentle, and 
humble person, who was particularly supportive of young people.

{\it Nils Lid Hjort:} I have many and uniformly warm remembrances 
of Peter. We had met and talked a few times at conferences, and then 
Peter invited me for a two-month stay in Canberra in 2000. 
This was both enjoyable, friendly, and fruitful. I remember 
fondly not only technical discussions and the free-flowing 
of ideas on blackboards (and since Peter could think twice 
as fast as anyone else, that somehow improved my own
arguing and thinking speed, or so I'd like to think), 
but also the positive, widely international, upbeat, but 
unstressed working atmosphere. Among the pluses for my Down Under
adventures were not merely meeting kangaroos in the wild 
while jogging and singing Schnittke, but teaming up with 
fellow visitors for several good projects, in particular with 
Gerda Claeskens; another sign of Peter's deep role in 
building partnerships and teams around him, by his sheer presence. 

Then Peter and Jeannie visited us in Oslo for a six-week 
period in the autumn of 2003. For their first day there, 
at least Jeannie was delighted that I had put on my Peter Hall 
t-shirt and that I gave him a {\it Hall of Fame} wristwatch. 
For these Oslo weeks he was therefore elaboratedly introduced 
at seminars and meetings as {\it Peter Hall of Fame};
everyone understood that all other Peter Halls were
considerably less famous.  
A couple of project ideas we developed together, in the middle
of Peter's dozens and dozens of other ongoing real-time 
projects, are still in my drawers and files, patiently
awaiting completion. Very few people can be as quietly 
and undramatically supremely efficient and productive 
as Peter. Luckily most of us others don't really have 
to, as long as we are doing decently well a decent 
proportion of the time. But once in a while, in my 
working life, when deadlines are approaching and I've 
lagged far behind, I put on my Peter Hall t-shirt, 
and think of him. It tends to work.

\begin{figure}[h]
\begin{center}
\hspace*{-.5cm} (a) \hspace*{5cm} (b) \\[.2cm]
\includegraphics[scale=.455]{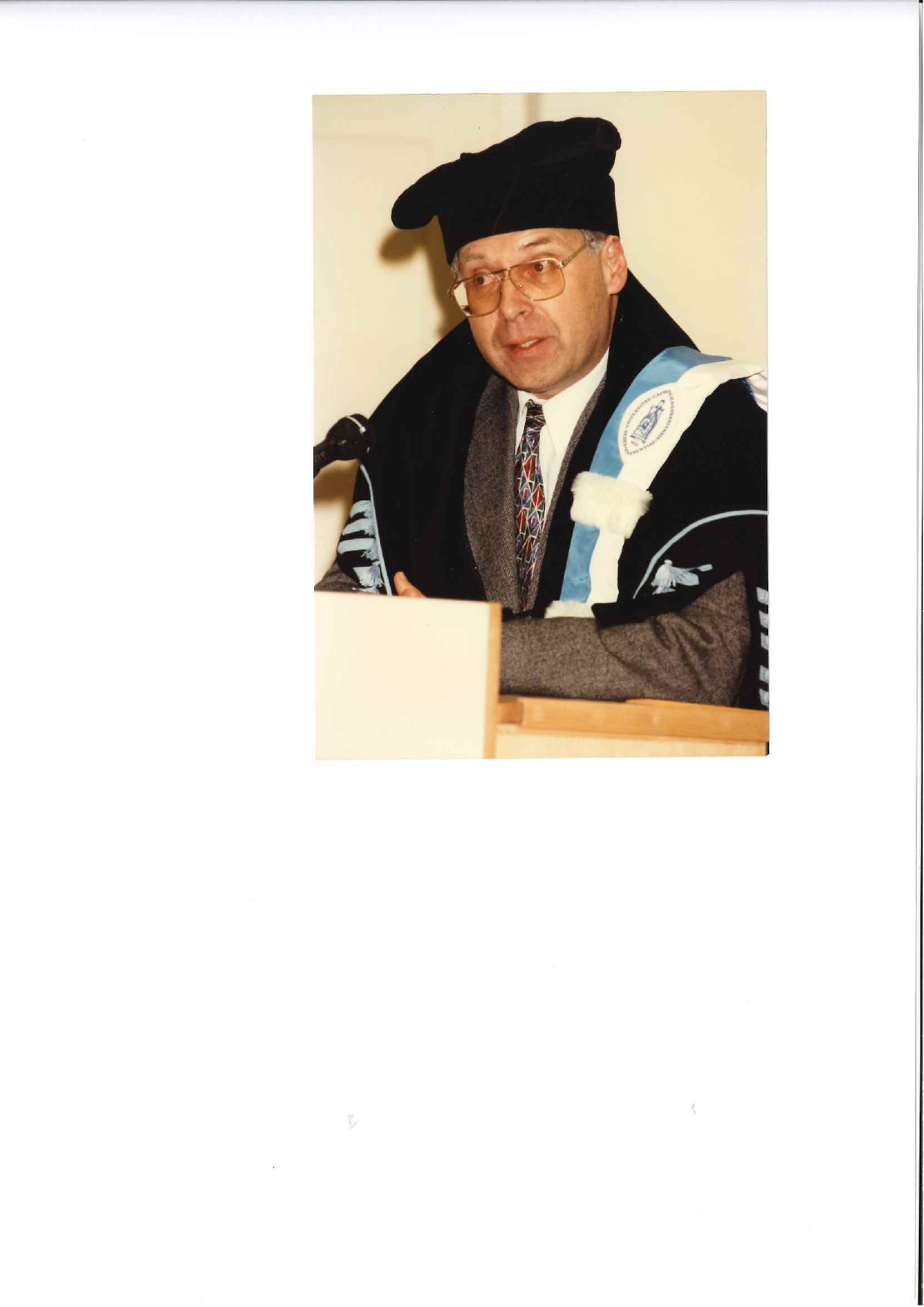} \ \ \ 
\includegraphics[scale=.42]{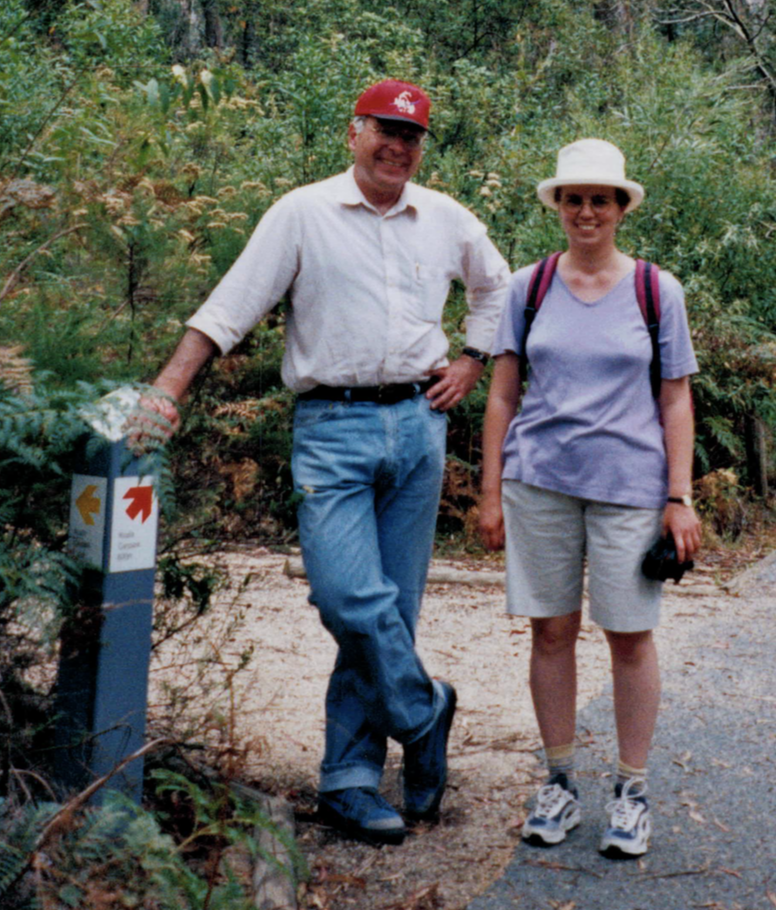} 
\end{center}
\vspace*{-.3cm}
\caption{(a) Peter at the occasion of his honorary 
doctorate at the Institute of Statistics in 
Louvain-la-Neuve in 1997; (b) Peter and Ingrid Van Keilegom 
in Tidbinbilla Nature Reserve near Canberra in 2002 
(picture taking by Jeannie Hall).}   
\label{figure:Peter}
\end{figure}

{\it Ingrid Van Keilegom:} 
I first met Peter in 1995 during one of Peter's many visits to 
Louvain-la-Neuve (LLN). At that time I was still a graduate 
student at Hasselt University. Two years later, in 1997, Peter 
obtained an honorary doctorate from the Institute of
Statistics in LLN (at the occasion of the fifth anniversary of the Institute), 
during which I discovered that Peter was not only 
a giant in his field but also a very human, modest, and kind person.   
Figure \ref{figure:Peter}(a) shows Peter at his acceptance speech. 
Later, in 2002, soon after I started working as a young faculty 
member in LLN, Peter invited me to Canberra for six weeks, 
a visit of which I have extremely positive memories.  
I am very grateful to Peter for having given me the opportunity 
to work with him there. During this visit Peter and I started 
working on two papers, and although Peter was very 
busy with many other things, it was difficult to stay on top 
of all new ideas and material that he was suggesting and 
adding to the papers, day after day. At some point during this 
visit Peter left Canberra for a 10-day visit to London, and I 
(naively) thought I could spend more time on broadening my 
knowledge on the two topics Peter had introduced to me. 
However, the next morning I received a fax of 20 pages of 
hand-written notes, containing a difficult proof that Peter 
had found during the flight to London. It took me the full 
next 10 days to unraffle all the details of the proof! 
Although Peter was very focused and busy with his work, 
he often took his visitors on a trip during the weekends.  
I enjoyed very much the trip to the Tidbinbilla Nature Reserve 
near Canberra, together with him and his wife Jeannie. A picture 
taken in this park by Jeannie is seen in Figure \ref{figure:Peter}(b).

After the visit to Canberra, Peter and I continued 
working on other projects and, in around 2004, Peter visited 
LLN for several weeks. I picked him up in the 
morning from the airport in Brussels.  He came
straight from Canberra and had been more or less 30 
hours underway. I supposed without asking that 
he would like to go to the hotel to take a rest.  
But when we were approaching the hotel, 
Peter insisted that I would drive immediately to 
the Institute in order to start working straight away. 
He spent the whole day at the Institute discussing 
with people and working in his office, before 
going finally to his hotel! I always wondered where 
he found the energy, the motivation and the strength 
to do this. He will be remembered by many of us 
as an extremely hard working person, and as an example to all of us.

\newpage 

\saveusalittle
\section {Introduction}
\label{section:intro}
\saveusalittle


For modelling data there are usually many options,
ranging from purely parametric, semiparametric, to fully 
nonparametric. There are also numerous ways 
in which to combine parametrics with nonparametrics, 
say estimating a model density by a combination
of a parametric fit with a nonparametric estimator,
or by taking a weighted average of parametric and 
nonparametric quantile estimators. 
This article concerns a proposal for a bridge
between a given parametric  model and a 
nonparametric likelihood-ratio method. 
We construct a hybrid likelihood function, based on 
the usual likelihood function for the
parametric model, say $L_n(\theta)$, with $n$
referring to sample size, and 
the empirical likelihood function for 
a given set of control parameters, say 
$R_n(\mu)$, where the $\mu$ parameters in question
are ``pushed through" the parametric model, 
leading to $R_n(\mu(\theta))$, say. Our hybrid
likelihood $H_n(\theta)$, defined in (\ref{eq:HEL}) below, 
is used for estimating the parameter vector of 
the working model; we term the $\hatt\theta_\hl$
in question the maximum hybrid likelihood estimator. 
This in turn leads to estimates of other 
quantities of interest. If $\psi$ is such a 
focus parameter, expressed via the working 
model as $\psi=\psi(\theta)$, then it is
estimated using $\hatt\psi_\hl=\psi(\hatt\theta_\hl)$. 

If the working parametric model is  correct,
these hybrid estimators lose a certain 
amount in terms of efficiency, when compared 
to the usual maximum likelihood estimator. 
We show, however, that the efficiency 
loss under ideal model conditions is typically 
a small one, and that the hybrid estimator
often  outperforms the maximum likelihood
when the working model is not correct. Thus 
the hybrid likelihood is seen to offer 
parametric robustness, or protection against model 
misspecification, by borrowing strength from the 
empirical likelihood, via the selected control parameters. 

Though our construction and methods can be 
lifted to e.g.~regression models, see 
Section \ref{subsection:regression} 
in the supplementary material, it is practical
to start with the simpler i.i.d.~framework, both
for conveying the basic ideas and for developing
theory. Thus, let $Y_1,...,Y_n$ be i.i.d.\
observations, stemming from some unknown density $f$. 
We wish to fit the data to some parametric family, 
say $f_\theta(y)=f(y,\theta)$,
with $\theta=(\theta_1,\ldots,\theta_p)^\tr \in \Theta$, 
where $\Theta$ is an open subset of $\mathbb{R}^p$ . 
The purpose of fitting the data to the model is 
typically to make inference about certain quantities 
$\psi=\psi(f)$, termed {\it focus parameters}, 
but without necessarily trusting the model fully. 
Our machinery for constructing 
robust 
estimators for these focus parameters 
involves certain extra parameters,
which we term {\it control parameters}, 
say $\mu_j=\mu_j(f)$ for $j=1,\ldots,q$. 
These are context-driven parameters, selected 
to safeguard against certain types of model
misspecification, and may or may not include 
the focus parameters. Suppose in general terms that 
$\mu=(\mu_1,\ldots,\mu_q)$ is identified 
via estimating equations,
$\E_f\,m_j(Y,\mu)=0$ for $j=1,\ldots,q$. 
Now consider 
\beq
\label{eq:basicEL}
R_n(\mu)=\max\Bigl\{\prodin (nw_i)\colon
   \sumin w_i=1,\sumin w_im(Y_i,\mu)=0,{\rm each\ }w_i>0\Bigr\}. 
\eeq  
This is the empirical likelihood function for $\mu$,
see \citet{Owen01}, with further discussions in 
e.g.~\citet{HMV09} and \citet[Ch.~11]{CLP16}. 
One might e.g.~choose 
$m(Y,\mu)=g(Y)-\mu$ for suitable $g=(g_1,\ldots,g_q)$, 
in which case the empirical likelihood
machinery gives confidence regions for the parameters 
$\mu_j=\E_f\,g_j(Y)$. 
We can now introduce the {\it hybrid likelihood (HL) function}  
\beq
\label{eq:HEL}
H_n(\theta)=L_n(\theta)^{1-a} R_n(\mu(\theta))^a, 
\eeq 
where $L_n(\theta) = \prod_{i=1}^n f(Y_i,\theta)$ is the ordinary likelihood,  
$R_n(\mu)$ is the empirical likelihood for $\mu$,
but here computed at the value $\mu(\theta)$, 
which is $\mu$ evaluated at $f_\theta$,  
and with $a$ being a balance parameter in $[0,1)$. 
The associated maximum HL 
estimator
is $\hatt\theta_\hl$, the maximiser of $H_n(\theta)$.  
If $\psi=\psi(f)$ is a parameter of interest, 
it is estimated as $\hatt\psi_\hl=\psi(f(\cdot,\hatt\theta_\hl))$.
This means first expressing $\psi$ in terms of the model 
parameters, say $\psi=\psi(f(\cdot,\theta))=\psi(\theta)$,  
and then plugging in the maximum HL 
estimator. 
The general approach (\ref{eq:HEL})
works for multidimensional vectors $Y_i$, so the
$g_j$ functions could e.g.~be set up to reflect
covariances. For one-dimensional cases, 
options include $m_j(Y,\mu_j)=I\{Y\le\mu_j\}-j/q$ 
($j=1,\ldots,q-1$) for quantile inference. 

The hybrid method (\ref{eq:HEL}) provides a bridge from the purely
parametric to the nonparametric empirical likelihood.
The $a$ parameter dictates the degree of balance. 
One can view (\ref{eq:HEL}) as a way for the empirical likelihood
to borrow strength from a parametric family, and, alternatively,
as a means of robustifying ordinary parametric model fitting
by incorporating precision control for one or more 
$\mu_j$ parameters. There might be practical circumstances
to assist one in selecting good $\mu_j$ parameters,
or good estimating equations, or these may be singled
out at the outset of the study as being quantities of 
primary interest. 

\saveusalittle 
\begin{example}
\label{example:example1}
\rm 
Let $f_\theta$ be the normal density with parameters $(\xi,\sigma^2)$, 
and take $m_j(y,\mu_j)=I\{y\le \mu_j\}-j/4$ for $j=1,2,3$. 
Then (1.1), with the ensuing 
$\mu_j(\xi,\sigma)=\xi+\sigma\,\Phi^{-1}(j/4)$ for $j=1,2,3$, 
can be seen as estimating the normal parameters 
in a way which modifies the parametric ML method
by taking into account the wish to have good model fit
for the three quartiles. Alternatively, it can be viewed
as a way of making inference for the three quartiles,
borrowing strength from the normal family in order  
to hopefully do better than simply using the three
empirical quartiles. 
\end{example}

\saveustwolittle 
\begin{example} 
\label{example:example2}
\rm 
Let $f_\theta$ be the Beta family with parameters $(b,c)$,
where ML estimates match moments for $\log Y$ and $\log(1-Y)$. 
Add to these the functions $m_j(y,\mu_j)=y^j-\mu_j$
for $j=1,2$. Again, this is Beta fitting with modification 
for getting the mean and variance about right, or moment
estimation borrowing strength from the Beta family. 
\end{example} 

\saveustwolittle 
\begin{example} 
\label{example:example4}
\rm 
Take your favourite parametric family $f(y,\theta)$,
and for an appropriate data set specify an interval
or region $A$ that actually matters. Then use 
$m(y,p)=I\{y\in A\}-p$ as the `control equation' above,
with $p=\Pr\{Y\in A\} = \int_A f(y,\theta) \, dy$. The effect is to push
the parametric ML estimates, softly or not softly depending
on the size of $a$, so as to make sure that the
empirical binomial estimate $\hatt p=n^{-1}\sumin I\{Y_i\in A\}$
is not far from the estimated 
$p(A,\hatt\theta)=\int_A f(y,\hatt\theta)\,\dd y$. 
This can also be extended to using a partition
of the sample space, say $A_1,\ldots,A_k$,
with control equations $m_j(y,p)=I\{y\in A_j\}-p_j$
for $j=1,\ldots,k-1$ (there is redundancy if trying
to include also $m_k$). It will be seen via our theory
 that the hybrid likelihood estimation strategy
in this case is large-sample equivalent to maximising 
$$(1-a)\ell_n(\theta)-\half an\,r(Q_n(\theta))
   =(1-a)\sumin\log f(Y_i,\theta)-\half an{Q_n(\theta)\over 1+Q_n(\theta)}, $$
where $r(w)=w/(1+w)$ and 
$Q_n(\theta)=\sum_{j=1}^k \{\hatt p_j-p_j(\theta)\}^2/\hatt p_j$. 
Here $\hatt p_j$ is the direct empirical binomial estimate
of $P\{Y\in A_j\}$ and $p_j(\hatt\theta)$ is
the model-based estimate. 
\end{example}

In Section \ref{section:basictheory} we explore the basic properties  
of HL estimators and the ensuing 
$\hatt\psi_\hl=\psi(\hatt\theta_\hl)$, under model 
conditions. Results here entail that the efficiency
loss is typically small, and of size $O(a^2)$ in terms 
of the balance parameter $a$. 
In Section \ref{section:outsidemodel} we study 
the behaviour of HL in $O(1/\rootn)$ neighbourhoods
of the parametric model.
It turns out that the HL estimator enjoys certain 
robustness properties, as compared to the ML estimator. 
Section \ref{section:finetuning} examines aspects
related to fine-tuning the balance parameter $a$ 
of (\ref{eq:HEL}), and we provide a recipe for 
its selection. An illustration of our HL methodology
is given in Section \ref{section:egypt}, 
involving fitting a Gamma model 
to data of Roman era Egyptian life-lengths, a century BC. 

Finally, coming back to the work of Peter Hall,  
a nice overview of all papers of Peter
on EL can be found in \citet{Changetal17}. We like to mention in 
particular the paper by \citet{Diciccioetal89}, in which the features and
behaviour of parametric and empirical likelihood functions 
are compared. We mention that Peter also
made very influential contributions to the somewhat related area of 
likelihood tilting, see e.g.~\citet{Choietal00}.

\saveusalittle
\section{Behaviour of HL under the parametric model}
\label{section:basictheory} 
\saveusalittle

The aim of this section is to explore asymptotic properties 
of the HL estimator under the parametric model 
$f(\cdot)=f(\cdot,\theta_0)$ for an appropriate true $\theta_0$.  
We establish the local asymptotic normality of 
HL, the asymptotic normality of the estimator 
$\hatt\theta_\hl$, and a version of the Wilks theorem. 
The HL estimator $\hatt\theta_\hl$ maximises 
\beq
\label{eq:logHn}
h_n(\theta)=\log H_n(\theta)
   =(1-a)\ell_n(\theta)+a\log R_n(\mu(\theta)) 
\eeq 
over $\theta$ (assumed here to be unique), where $\ell_n(\theta) = \log L_n(\theta)$. 
We need to analyse the local behaviour 
of the two parts of $h_n(\cdot)$.  

Consider the localised empirical likelihood 
$R_n(\mu(\theta_0 + s/\rootn))$, where $s$ belongs to some 
arbitrary compact $S\subset \mathbb{R}^p$. For simplicity 
we write 
$m_{i,n}(s)=m(Y_i,\mu(\theta_0 + s/\rootn))$. 
Also, consider the functions 
$G_n(\lambda,s)=\sumin 2\log\{1+\lambda^\tr m_{i,n}(s)/\rootn\}$ and
$G_n^*(\lambda,s)=2\lambda^\tr V_n(s)-\lambda^\tr W_n(s)\lambda$ 
of the $q$-dimensional $\lambda$, where 
$V_n(s)=n^{-1/2}\sumin m_{i,n}(s)$ and 
$W_n(s)=n^{-1}\sumin m_{i,n}(s)m_{i,n}(s)^\tr$. 
Hence $G_n^*$ is the two-term Taylor expansion of $G_n$.  

We now re-express the EL statistic in terms of Lagrange multipliers 
$\hatt\lambda_n$, which is pure analysis, not yet having anything 
to do with random variables, per se:
$-2\log R_n(\mu(\theta_0 + s/\rootn))=\max_\lambda 
G_n(\lambda,s)=G_n(\hatt\lambda_n(s),s)$, 
with $\hatt\lambda_n(s)$ 
the solution to 
$\sumin m_{i,n}(s)/\{1+\lambda^\tr m_{i,n}(s)/\rootn\}=0$ for all $s$. 
This basic translation from the EL 
definition via Lagrange multipliers is contained in 
\citet[Ch.~11]{Owen01}; for a detailed proof, along 
with further discussion, see \citet[Remark 2.7]{HMV09}. 
The following lemma
is crucial for understanding the basic properties of HL.  
The proof is in Section \ref{subsection:prooflemma1} 
in the supplementary material. For any matrix 
$A=(a_{j,k})$, $\|A\| = (\sum_{j,k} a_{j,k}^2)^{1/2}$ denotes the 
Euclidean norm. 

\saveustwolittle 
\begin{lemma}
\label{lemma:lemma1}
For a compact $S\subset \mathbb{R}^p$, 
suppose that (i) $\sup_{s \in S} \|V_n(s)\| = O_\pr(1)$; 
(ii) $\sup_{s \in S}\|W_n(s)-W\| \arr_\pr 0$, where 
$W = \Var\,m(Y,\mu(\theta_0))$ is of full rank;
(iii) $n^{-1/2} \sup_{s \in S} \maxin\|m_{i,n}(s)\|\allowbreak\arr_\pr0$. 
Then, the maximisers $\hatt\lambda_n(s)=\argmax_\lambda G_n(\lambda,s)$
and $\lambda_n^*(s)=\argmax_\lambda G_n^*(\lambda,s)=W_n^{-1}(s)V_n(s)$ 
are both $O_\pr(1)$ uniformly in $s \in S$, and 
$ \sup_{s \in S} |\max_\lambda G_n(\lambda,s)-\max_\lambda G_n^*(\lambda,s)| 
= \sup_{s \in S} |G_n(\hatt\lambda_n(s),s)-G_n^*(\lambda_n^*(s),s)| \allowbreak \arr_\pr 0$.  
\end{lemma}

\saveustwolittle 
Note that we have an explicit expression for the maximiser
of $G_n^*(\cdot,s)$, hence also its maximum, $\allowbreak$
$\max_\lambda G_n^*(\lambda,s)=V_n(s)^\tr W_n^{-1}(s)V_n(s)$.
It follows that in situations covered by Lemma \ref{lemma:lemma1}, 
$-2\log R_n(\mu(\theta_0+s/\rootn))=V_n(s)^\tr W_n^{-1}(s)V_n(s)+o_\pr(1)$,
uniformly in $s\in S$.  Also,  by the Law of Large Numbers, condition
(ii) of Lemma \ref{lemma:lemma1} is valid if $\sup_s \|W_n(s)-W_n(0)\| \arr_\pr 0$.
If $m$ and $\mu$ are smooth, 
then the latter holds using the Mean Value Theorem.  
For the quantile example, Example \ref{example:example1}, 
we can use results on the oscillation behaviour 
of empirical distributions (see \citet{Stute1982}).

For our Theorem \ref{theorem:theorem1} below
we need assumptions on the $m(y,\mu)$ function
involved in (\ref{eq:basicEL}), 
and also on how $\mu=\mu(f_\theta)=\mu(\theta)$  behaves 
close to $\theta_0$. In addition to  
$\E\,m(Y,\mu(\theta_0))=0$, 
we assume that 
\beq
\label{eq:msmooth1}
\sup_{s \in S} \|V_n(s)-V_n(0) - \xi_n s \| = o_\pr(1), 
\eeq 
with $\xi_n$ of dimension $q\times p$ 
tending in probability to $\xi_0$.
Suppose for illustration 
that $m(y,\mu(\theta))$ has a derivative at $\theta_0$, and write 
$m(y,\mu(\theta_0+\eps))=m(y,\mu(\theta_0))+\xi(y)\eps+r(y,\eps)$, 
for the appropriate $\xi(y)=\dell m(y,\mu(\theta_0))/\dell\theta$,
a $q\times p$ matrix, 
and with a remainder term $r(y,\eps)$. This fits
with (\ref{eq:msmooth1}), with 
$\xi_n=n^{-1}\sumin\xi(Y_i)\arr_\pr\xi_0=\E\,\xi(Y)$,
as long as $n^{-1/2}\sumin r(Y_i,\allowbreak s/\rootn)\arr_\pr 0$ 
uniformly in $s$. In smooth cases we would typically have 
$r(y,\eps)=O(\|\eps\|^2)$, making the mentioned term 
of size $O_\pr(1/\rootn)$.  On the other hand, 
when $m(y,\mu(\theta))=I\{y\le \mu(\theta)\}-\alpha$,
we have $V_n(s)-V_n(0) = f(\mu(\theta_0),\theta_0) s + O_\pr(n^{-1/4})$ 
uniformly in $s$ (see \citet{Stute1982}).  

We rewrite the log-HL in terms of a local 
$1/\rootn$-scale perturbation around $\theta_0$:
\beq
\label{eq:Anfunction}
\begin{array}{rcl}
A_n(s)&=&h_n(\theta_0+s/\rootn)-h_n(\theta_0) \\
   &=&(1-a)\{\ell_n(\theta_0+s/\rootn)-\ell_n(\theta_0)\} 
+a\{\log R_n(\mu(\theta_0+s/\rootn))-\log R_n(\mu(\theta_0))\}. 
\end{array}
\eeq 
Below we show that $A_n(s)$ converges weakly to a 
quadratic limit $A(s)$, uniformly in $s$ over compacta, 
which then leads to our most important insights 
concerning HL-based estimation and inference. 
By the multivariate Central Limit Theorem,
\beq
\label{eq:CLTatwork}
\begin{pmatrix} U_{n,0} \\ V_{n,0} \end{pmatrix}
=\begin{pmatrix} n^{-1/2}\sumin u(Y_i,\theta_0) \\ 
   n^{-1/2}\sumin m(Y_i,\mu(\theta_0)) \end{pmatrix}
\arr_d 
\begin{pmatrix} U_0 \\ V_0 \end{pmatrix}
   \sim\N_{p+q}(0,\Sigma), \ \ \ \mbox{where } 
   \Sigma = \begin{pmatrix}J & C \\ C^\tr & W\end{pmatrix}. 
\eeq 
Here, $u(y,\theta) = \partial \log f(y,\theta)/\partial \theta$ 
is the score function, $J=\Var\,u(Y,\theta_0)$
is the Fisher information matrix of dimension $p \times p$, 
$C=\E\,u(Y,\theta_0) m(Y,\mu(\theta_0))^\tr$ is of dimension $p\times q$,
and $W=\Var\,m(Y,\mu(\theta_0))$ as before.
The $(p+q)\times(p+q)$ variance matrix 
$\Sigma$ is assumed to be positive definite. 
This ensures that the parametric and empirical likelihoods
do not ``tread on one another's toes'',  
i.e.~that the $m_j(y,\mu(\theta))$ functions are not in the span
of the score functions, and vice versa. 

\saveusalittle 
\begin{theorem}
\label{theorem:theorem1}
Suppose that smootheness conditions on $\log f(y,\theta)$ hold,
as spelled out in Section \ref{subsection:prooftheorem1}; 
the conditions of Lemma \ref{lemma:lemma1} are in force,
along with condition (\ref{eq:msmooth1}) with the appropriate $\xi_0$,
for each compact $S\subset \mathbb{R}^p$;  
and that $\Sigma$ has full rank. Then, for each compact $S$, 
$A_n(s)$ converges weakly to 
$A(s)=s^\tr U^*-\half s^\tr J^*s$,  
in the function space ${\ell}^\infty(S)$ endowed with the uniform topology, 
where 
$U^*=(1-a)U_0-a\xi_0^\tr W^{-1}V_0$ and 
$J^*=(1-a)J+a\xi_0^\tr W^{-1}\xi_0$. 
Here $U^*\sim\N_p(0,K^*)$, with variance matrix 
$K^*=(1-a)^2J + a^2\xi_0^\tr W^{-1}\xi_0
   -a(1-a)(CW^{-1}\xi_0+\xi_0^\tr W^{-1}C^\tr)$. 
\end{theorem}

\saveusalittle 
The theorem, proved in Section \ref{subsection:prooftheorem1}
of the supplementary material, is valid for each fixed 
balance parameter $a$ in (\ref{eq:HEL}),
with $J^*$ and $K^*$ also depending on $a$. 
We discuss ways of fine-tuning $a$ 
in Section \ref{section:finetuning}. 

The $p\times q$-dimensional block component $C$ 
of the variance matrix $\Sigma$ of (\ref{eq:CLTatwork}) 
can be worked with and represented in different ways. Suppose 
that $\mu$ is differentiable at $\theta=\theta_0$, and
denote the vector of partial derivatives by 
$\dellmudelltheta$, with derivatives at $\theta_0$, 
and with this matrix arranged as a $p\times q$ matrix,
with columns $\dell\mu_j(\theta_0)/\dell\theta$ for $j=1,\ldots,q$.
From $\int m(y,\mu(\theta))f(y,\theta)\,\dd y=0$ for all $\theta$
follows the $q\times p$-dimensional equation 
$\int m^*(y,\mu(\theta_0))f(y,\theta_0)\,\dd y\,(\dellmudelltheta)^\tr
   +\int m(y,\mu(\theta_0))f(y,\theta_0)u(y,\theta_0)^\tr\,\dd y=0$, 
where $m^*(y,\mu)=\dell m(y,\mu)/\dell\mu$, 
in case $m$ is differentiable with respect to $\mu$.
This means $C=-\dellmudelltheta\,\E_\theta\,m^*(Y,\mu(\theta_0))$.
If $m(y,\mu)=g(y)-\mu$, for example, corresponding 
to parameters $\mu=\E\,g(Y)$, we have $C=\dellmudelltheta$. 
Also, using (\ref{eq:msmooth1}) 
we have $\xi_0=-(\dellmudelltheta)^\tr$,
of dimension $q\times p$. Applying Theorem \ref{theorem:theorem1}
yields $U^*=(1-a)U_0+a\dellmudelltheta W^{-1}V_0$, along with 
\vspace{-0.5cm}
\beq
\label{eq:JstarKstar}
J^*=(1-a)J+a\dellmudelltheta W^{-1}\Bigl(\dellmudelltheta\Bigr)^\tr
   \quadandquad 
K^*=(1-a)^2J+\{1-(1-a)^2\}\dellmudelltheta W^{-1}\Bigl(\dellmudelltheta\Bigr)^\tr. 
\eeq 

For the following corollary of Theorem \ref{theorem:theorem1}, 
we need to introduce the random function 
$\Gamma_n(\theta) = n^{-1}\{h_n(\theta)- h_n(\theta_0)\}$ 
along with its population version
\beq
\label{eq:Gammafunction}
\Gamma(\theta) = -(1-a)\,\KL(f_{\theta_0},f_\theta)
- a\,\E\,\log\big(1+\xi(\theta)^\tr m(Y,\mu(\theta))\big),
\eeq 
with $\hatt\theta_\hl$ as the argmax of $\Gamma_n(\cdot)$. 
Here $\KL(f,f_\theta)=\int f\log (f/f_\theta)\,\dd y$ is 
the Kullback--Leibler divergence, in this case from 
$f_{\theta_0}$ to $f_\theta$, and with $\xi(\theta)$ the solution of
$\E\,m(Y,\mu(\theta))/\{1+\xi^\tr m(Y,\mu(\theta)\}= 0$
(that this solution exists and is unique is a consequence 
of the proof of Corollary \ref{cor:cor1} below). 

\saveusalittle 
\begin{corollary}
\label{cor:cor1}
Under the conditions of Theorem \ref{theorem:theorem1} and under 
conditions (A1)--(A3) given in Section \ref{subsection:proofcor1} 
of the supplementary material, 
(i) there is consistency of $\hatt\theta_\hl$ towards $\theta_0$; 
(ii) $\rootn(\hatt\theta_\hl-\theta_0)\arr_d
   (J^*)^{-1}U^*\sim\N_p\big(0,(J^*)^{-1}K^*(J^*)^{-1}\big)$; 
and (iii) 
$2\{h_n(\hatt\theta_\hl)-h_n(\theta_0)\}\arr_d(U^*)^\tr (J^*)^{-1}U^*$.
\end{corollary}

\saveusalittle 
These results allow  us to construct confidence
regions for $\theta_0$ and confidence intervals 
for its components. Of course we are not merely interested 
in the individual parameters of a model, but in 
certain functions of them, namely focus parameters.
Assume $\psi=\psi(\theta)=\psi(\theta_1,\ldots,\theta_p)$
is such a parameter, with $\psi$  differentiable at $\theta_0$ and denote 
$c=\dell\psi(\theta_0)/\dell\theta$. The HL estimator
for this $\psi$ is the plug-in $\hatt\psi_\hl=\psi(\hatt\theta_\hl)$. 
With $\psi_0=\psi(\theta_0)$ as the true parameter value, 
we then have via the delta method that
\begin{equation}
\label{cor:cor3}
\rootn(\hatt\psi_\hl-\psi_0)\arr_d c^\tr (J^*)^{-1}U^* \sim\N(0,\kappa^2),
\ \ \mbox{where} \ \kappa^2=c^\tr(J^*)^{-1}K^*(J^*)^{-1}c.\end{equation}
The focus parameter $\psi$ could, for example, be
one of the components of  $\mu=\mu(\theta)$ 
used in the EL part of the HL, say
$\mu_j$, for which 
$\rootn(\hatt\mu_{j,\hl}-\mu_{0,j})$ has a normal 
limit with variance 
$({\dell\mu_j\over \dell\theta})^\tr(J^*)^{-1}K^*(J^*)^{-1}
{\dell\mu_j\over \dell\theta}$, 
in terms of 
${\dell\mu_j\over \dell\theta}=\dell\mu_j(\theta_0)/\dell\theta$. 
Armed  with Corollary \ref{cor:cor1}, 
we can set up Wald and likelihood-ratio type confidence 
regions and tests for $\theta$, and confidence intervals for $\psi$. 
Consistent estimators $\hatt J^*$ 
and $\hatt K^*$ of 
$J^*$ and $K^*$ would then be required,  
but these are readily obtained via plug-in. 
Also, an estimate of $J^*$ is typically obtained
via the Hessian of the optimisation algorithm 
used to find the HL estimator in the first place. 

In order to investigate how much is lost in efficiency 
when using the HL estimator under model conditions, 
consider the case of small $a$. We have 
$J^*=J+aA_1$ and $K^*=J+aA_2+O(a^2)$, 
with $A_1=\xi_0^\tr W^{-1}\xi_0-J$ and $A_2=-2J-CW^{-1}\xi_0-\xi_0^\tr W^{-1}C^\tr$. 
For the class of functions of the form 
$m(y,\mu)=g(y)-T(\mu)$, corresponding to $\mu=T^{-1}(\E\,g(Y))$, 
we have $A_2=2A_1$. It is assumed that $T(\cdot)$ has 
a continuous inverse at $\mu(\theta)$ for $\theta$
in a neighbourhood of $\theta_0$. Writing $(J^*)^{-1}K^*(J^*)^{-1}$
as $(J^{-1}-aJ^{-1}A_1J^{-1})(J+aA_2)(J^{-1}-aJ^{-1}A_1J^{-1})+O(a^2)$,
therefore, one finds that this is $J^{-1}+O(a^2)$,  
which in particular means that the efficiency loss 
is very small  when $a$ is small. 

\saveusalittle
\section{Hybrid likelihood outside model conditions}
\label{section:outsidemodel}
\saveusalittle 

In Section \ref{section:basictheory} we investigated   the hybrid 
likelihood estimation strategy under 
the conditions of the parametric model. 
Under suitable conditions, the HL is consistent and asymptotically
normal, with a certain mild loss of efficiency 
under model conditions, compared to the 
parametric 
ML method, the special case $a=0$. 
In the present section we  investigate the 
behaviour of the HL outside the conditions
of the parametric  model, which is now viewed as a working model. 
It turns out that  HL  often outperforms 
 ML  by reducing model bias, 
which in mean squared error terms might more than 
compensate for a slight increase in variability. 
This in turn calls for methods for fine-tuning
the balance parameter $a$ in our basic 
hybrid construction (\ref{eq:HEL}), and we shall
deal with this problem too, in Section \ref{section:finetuning}. 

Our framework for investigating such properties
involves extending the working model $f(y,\theta)$ 
to a $f(y,\theta,\gamma)$ model, where 
$\gamma=(\gamma_1,\ldots,\gamma_r)$ is a vector 
of extra parameters. There is a null value $\gamma=\gamma_0$
which brings this extended model back to the working model. 
We now examine behaviour of the ML and the HL 
schemes when $\gamma$ is in the neighbourhood of $\gamma_0$. 
Suppose in fact that 
\beq
\label{eq:ftruewithdelta}
f_\true(y)=f(y,\theta_0,\gamma_0+\delta/\rootn),
\eeq 
with the $\delta=(\delta_1,\ldots,\delta_r)$ parameter
dictating the relative distance from the null model. 
In this framework, suppose an estimator $\hatt\theta$ 
has the property that 
\vspace{-0.3cm}
\beq
\label{eq:basic1}
\rootn(\hatt\theta-\theta_0)\arr_d \N_p(B\delta,\Omega), 
\eeq 
with a suitable $p\times r$ matrix $B$ related to 
how much the model bias affects the estimator of $\theta$,
and limit variance matrix $\Omega$. Then a parameter 
$\psi=\psi(f)$ of interest can in this wider framework 
be expressed as $\psi=\psi(\theta,\gamma)$,
with true value $\psi_\true=\psi(\theta_0,\gamma_0+\delta/\rootn)$. 
The spirit of these investigations is that the statistician
uses the working model with only $\theta$ present, 
without knowing the extension model or the size
of the $\delta$ discrepancy. 
The ensuing estimator
for $\psi$ is hence $\hatt\psi=\psi(\hatt\theta,\gamma_0)$.
The delta method 
then leads to 
\beq
\label{eq:basic2}
\rootn(\hatt\psi-\psi_\true)\arr_d\N(b^\tr\delta,\tau^2), 
\eeq 
with $b=B^\tr\dellone-\delltwo$ and 
$\tau^2=(\dellone)^\tr\Omega\dellone$, and with 
partial derivatives evaluated at the working model, i.e.\ at
$(\theta_0,\gamma_0)$. 
The limiting mean squared error, for such an estimator of $\mu$, is 
$\mse(\delta)=(b^\tr\delta)^2+\tau^2$. 
Among the consequences of using the narrow working model 
when it is moderately wrong, at the level of 
$\gamma=\gamma_0+\delta/\rootn$, is the bias $b^\tr\delta$. 
The size of this bias depends on the focus
parameter, and  it may be zero for some foci,
even when the model is incorrect.

We  now examine both the ML and the HL methods
in this framework, exhibiting the associated 
$B$ and $\Omega$ matrices and hence the mean 
squared errors, via (\ref{eq:basic2}). 
Consider the parametric ML estimator $\hatt\theta_\pl$ first. 
To present the necessary results, consider the 
$(p+r)\times(p+r)$ Fisher information matrix
\beq
\label{eq:fisher}
J_\wide=\begin{pmatrix} J_{00} &J_{01} \\ J_{10} &J_{11} \end{pmatrix} 
\eeq 
for the $f(y,\theta,\gamma)$ model, computed at 
the null values $(\theta_0,\gamma_0)$. 
In particular, the $p\times p$ block $J_{00}$,
corresponding to the model with only $\theta$ and without $\gamma$,
is equal to the earlier $J$ matrix of (\ref{eq:CLTatwork})
and appearing in Theorem~\ref{theorem:theorem1} etc. 
Here one can demonstrate, under appropriate 
mild regularity conditions, that 
$ \rootn(\hatt\theta_\narr-\theta_0)
   \arr_d\N_p(J_{00}^{-1}J_{01}\delta,J_{00}^{-1})$. 
Just as (\ref{eq:basic2}) followed from (\ref{eq:basic1}), 
one finds for $\hatt\psi_\pl=\psi(\hatt\theta_\pl)$ that 
\beq
\label{eq:muhatnarrlimit}
\rootn(\hatt\psi_\pl-\psi_\true)\arr_d
   \N(\omega^\tr\delta,\tau_0^2), 
\eeq
featuring $\omega=J_{10}J_{00}^{-1}\dellone-\delltwo$ 
and $\tau_0^2=(\dellone)^\tr J_{00}^{-1}\dellone$.
See also \citet{HjortClaeskens03} and 
\citet[Chs.~6, 7]{ClaeskensHjort08} for further details,
discussion, and precise regularity conditions.  

In such a situation, with a clear interest parameter $\psi$, 
we use the HL to get $\hatt\psi_\hl=\psi(\hatt\theta_\hl,\gamma_0)$.  
We work out what happens with $\hatt\theta_\hl$ 
in this framework, generalising what is found in the previous section. 
Introduce $S(y)=\dell\log f(y,\theta_0,\gamma_0)/\dell\gamma$, 
the score function 
in direction of these extension parameters, and let 
$K_{01}=\int f(y,\theta_0)m(y,\mu(\theta_0))S(y)\,\dd y$, 
of dimension $q\times r$, along with 
$L_{01}=(1-a)J_{01}-a(\dellone)^\tr W^{-1}K_{01}$,
of dimension $p\times r$, and with transpose $L_{10}=L_{01}^\tr$. 
The following is proved in Section \ref{subsection:prooftheorem2}. 

\vspace{-0.37cm}
\begin{theorem}
\label{theorem:theorem2}
Assume data stem from the extended $p+r$-dimensional
model (\ref{eq:ftruewithdelta}), and that the conditions
listed in Corollary \ref{cor:cor1} are in force. 
For the HL method, with the focus parameter $\psi=\psi(f)$
built into the construction (\ref{eq:HEL}), 
results (\ref{eq:basic1})--(\ref{eq:basic2}) hold, with 
$B=(J^*)^{-1}L_{01}$ and 
$\Omega=(J^*)^{-1}K^*(J^*)^{-1}$. 
\end{theorem}

\vspace{-0.65cm}
The limiting distribution for $\hatt\psi_\hl=\psi(\hatt\theta_\hl)$ 
can again be read off, just as 
(\ref{eq:basic2}) follows from (\ref{eq:basic1}):  
\beq
\label{eq:hatmuHLlimit}
\rootn(\hatt\psi_\hl-\psi_\true)\arr_d\N(\omega_\hl^\tr\delta,
   \tau_{0,\hl}^2), 
\eeq 
with $\omega_\hl=L_{10}(J^*)^{-1}\dellone-\delltwo$ 
and $\tau_{0,\hl}^2=(\dellone)^\tr (J^*)^{-1}K^*(J^*)^{-1}\dellone$. 
The quantities involved in these large-sample properties of the HL estimator 
depend on the balance parameter $a$ employed in 
the basic HL construction (\ref{eq:HEL}). 
For $a=0$ we are back to the ML, with (\ref{eq:hatmuHLlimit})
specialising to (\ref{eq:muhatnarrlimit}).
As $a$ moves away from zero, more emphasis is placed
on the EL part, in effect pushing
$\theta$ so $n^{-1}\sumin m(Y_i,\mu(\theta))$ 
gets closer to zero. The result is typically a lower bias 
$|\omega_\hl(a)^\tr\delta|$, compared to $|\omega^\tr\delta|$,
and a slightly larger standard deviation $\tau_{0,\hl}$, 
compared to $\tau_0$. 
Thus selecting a good value of $a$ is a bias-variance
balancing game, which we discuss in the following section.

\saveustwolittle 
\section{Fine-tuning the balance parameter}
\label{section:finetuning}

\saveustwolittle 
The basic HL construction of (\ref{eq:HEL}) 
first entails selecting context relevant control
parameters $\mu$, 
and then a focus parameter $\psi$. 
A special case is that of using the focus $\psi$ as the single control
parameter. In each case, there is also the balance parameter $a$
to decide upon. Ways of fine-tuning the balance 
are discussed 
here. 

{\bf Balancing robustness and efficiency.} 
By allowing the empirical likelihood to be combined
with the likelihood from a given parametric model, 
one may buy robustness, via the control parameters $\mu$
in the HL construction, at the expense of a certain
mild loss of efficiency. One way to fine-tune the 
balance, after having decided on the control parameters,  
is to select $a$ so that the loss of efficiency 
under the conditions of the parametric working model 
is limited by a fixed, small amount, say 5\%. 
This may be achieved by using the corollaries of 
Section \ref{section:basictheory}  
by comparing 
the inverse Fisher information matrix $J^{-1}$,
associated with the ML estimator, to the sandwich matrix 
$(J^*_a)^{-1}K^*_a(J^*_a)^{-1}$, for the HL estimator.
Here we refer to the corollaries of 
Section \ref{section:basictheory}, see e.g.~(\ref{eq:JstarKstar}), 
and have added the subscript $a$, for emphasis. 
If there is special interest in some focus parameter 
$\psi$, one may select $a$ so that 
\beq
\label{eq:efficiencycontrol}
\kappa_a=\{c^\tr(J^*_a)^{-1}K^*_a(J^*_a)^{-1}c\}^{1/2} 
   \le(1+\eps)\kappa_0=(1+\eps)(c^\tr J^{-1}c)^{1/2}, 
\eeq 
with $\eps$ the required threshold. With $\eps=0.05$,
for example, one ensures that confidence intervals are 
only 5\% wider than those based on the ML, but with 
the additional security of having controlled well for 
the $\mu$ parameters in the process, e.g.~for robustness reasons. 
Pedantically speaking, in (\ref{cor:cor3}) 
there is really a $c_a=\dell\psi(\theta_{0,a})/\dell\theta$
also depending on the $a$, associated with the limit in probability
$\theta_{0,a}$ of the HL estimator, but when 
discussing efficiencies at the parametric model,
the $\theta_{0,a}$ is the same as the true 
$\theta_0$, so $c_a$ is the same as $c=\dell\psi(\theta_0)/\dell\theta$. 
A concrete illustration of this approach is in the 
following section. 

{\bf Features of the mse(a).}
The methods above, as with (\ref{eq:efficiencycontrol}),
rely on the theory developed in Section \ref{section:basictheory},
under the conditions of the parametric working model.
In what follows we need the 
theory given in 
Section \ref{section:outsidemodel}, 
examining the behaviour
of the HL estimator in a neighbourhood around the 
working model. Results there can first 
be used to examine the limiting mse 
properties for the ML and the HL estimators 
where it will be seen that the HL often can behave
better; a slightly larger variance is being compensated
for with a smaller modelling bias. Secondly,
the mean squared error curve, as a function of 
the balance parameter $a$, can be estimated from 
data. This leads to the idea of selecting $a$
to be the minimiser of this estimated risk curve, pursued below. 
 

For a given focus parameter $\psi$, the limit mse
when using the HL with parameter $a$ is found from 
(\ref{eq:hatmuHLlimit}):  
\beq
\label{eq:mseofa}
\mse(a)=\{\omega_\hl(a)^\tr\delta\}^2+\tau_{0,\hl}(a)^2.
\eeq 
The first exercise is to evaluate this curve, 
as a function of the balance parameter $a$, in situations 
with given model extension parameter $\delta$. 
The $\mse(a)$ at $a=0$ corresponds to the mse 
for the ML estimator. If $\mse(a)$ is smaller than $\mse(0)$, 
for some $a$, then the HL is doing a better job than the ML. 

\begin{figure}[h] 
\begin{center}
\vspace*{-.3cm}
\hspace*{.5cm} (a) \hspace*{7cm} (b) \\[-.3cm]
\includegraphics[width=7cm,height=6cm]{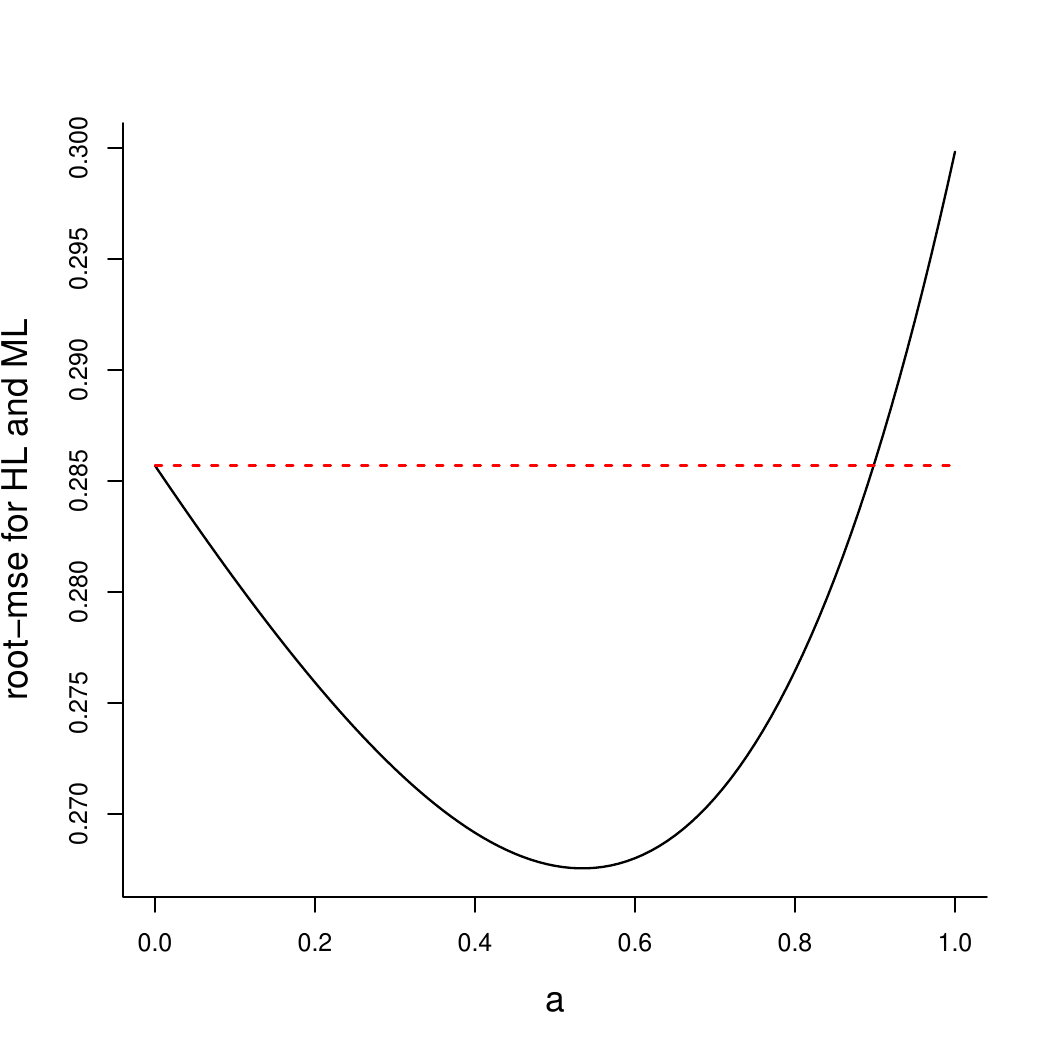} \ \ \ \ \ \includegraphics[width=7cm,height=6cm]{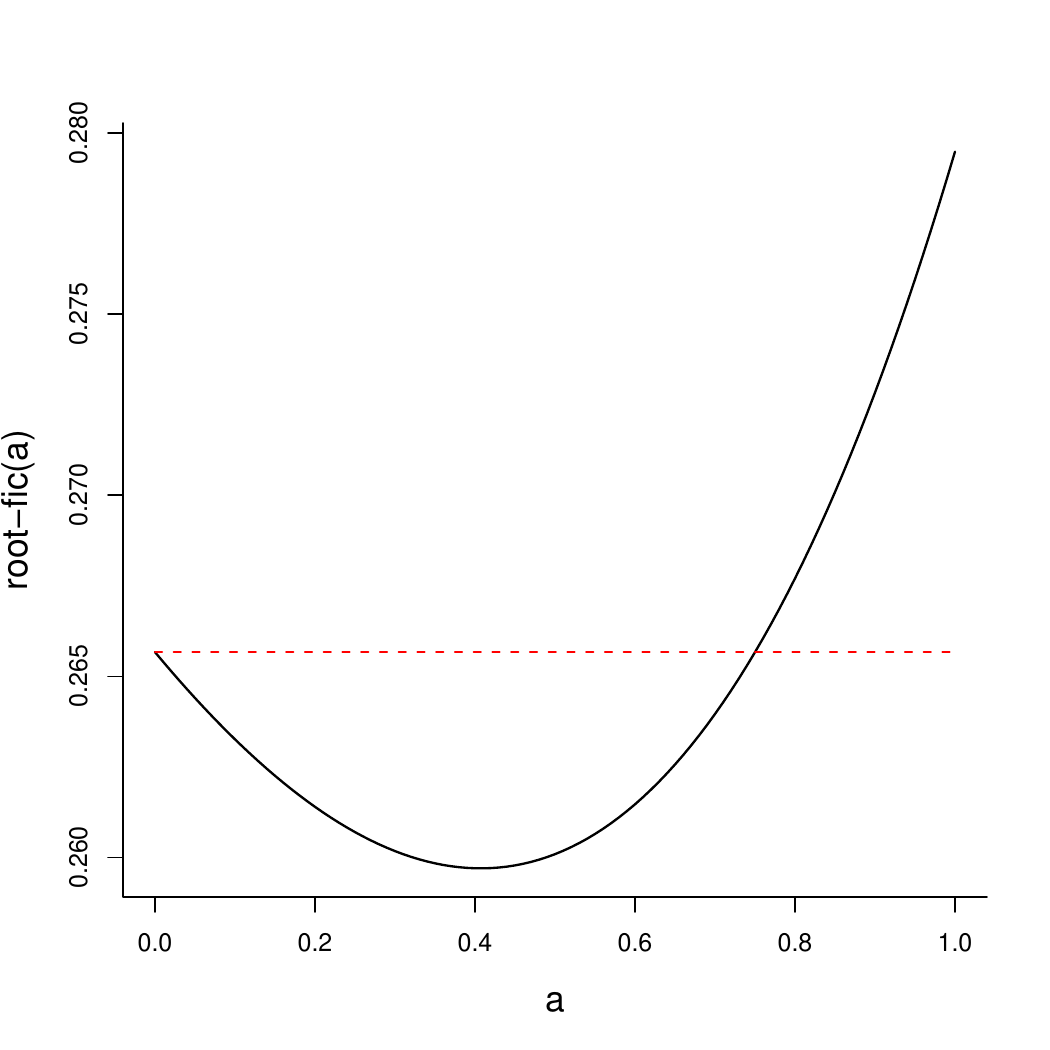}
\end{center}
\vspace*{-.3cm}
\caption{(a) The dotted horizontal line indicates the
root-mse for the ML estimator, and the full curve
the root-mse for the HL estimator, as a function
of the balance parameter $a$ in the HL construction.
(b) The root-$\fic(a)$,
as a function of the 
balance parameter $a$, 
constructed on the basis
of $n=100$ simulated observations, from a case 
where $\gamma=1+\delta/\rootn$, with $\delta$ described
in the text.}  
\label{figure:hel2}
\end{figure}

\saveustwolittle 
Figure \ref{figure:hel2}(a) displays the root-$\mse(a)$
curve in a simple setup, where the parametric start model
is $\Beta(\theta,1)$, with density 
$\theta y^{\theta-1}$, and the focus parameter used 
for the HL construction is $\psi=\E\,Y^2$, which 
is $\theta/(\theta+2)$ under model conditions.
The extended model, under which we examine the mse 
properties of the ML and the HL, is the $\Beta(\theta,\gamma)$,
with $\gamma=1+\delta/\rootn$ in (\ref{eq:ftruewithdelta}). 
The $\delta$ for this illustration is chosen to be 
$Q^{1/2}=(J^{11})^{1/2}$, from (\ref{eq:DntoD}) below,
which may be interpreted as one standard deviation away 
from the null model. The root-$\mse(a)$ curve, 
computed via numerical integration, shows that the HL estimator 
$\hatt\theta_\hl/(\hatt\theta_\hl+2)$ does better
than the parametric ML estimator $\hatt\theta_\pl/(\hatt\theta_\pl+2)$, 
unless $a$ is close to 1. Similar curves are seen for 
other $\delta$, for other focus parameters, and 
for more complex models. Occasionally, $\mse(a)$ 
is increasing in $a$, indicating in such cases that 
ML is better than HL, but this
typically happens only when the model discrepancy
parameter $\delta$ is small, i.e.~when the working
model is nearly correct. 

It is of interest to note that $\omega_\hl(a)$ 
in (\ref{eq:hatmuHLlimit})  
starts out for $a=0$ at $\omega=J_{10}J_{00}^{-1}\dellone-\delltwo$
in (\ref{eq:muhatnarrlimit}), associated with the 
ML method, 
but then it decreases in size
towards zero, as $a$ grows from zero to one. 
Hence, when HL employs only a small part of the 
ordinary log-likelihood in its construction, 
the consequent $\hatt\psi_{\hl,a}$ has small bias, 
but potentially a bigger variance than ML. 
The HL may thus be seen as a debiasing operation,
for the control and focus parameters, in cases
where the parametric 
model $f(\cdot,\theta)$
cannot be fully trusted.



{\bf Estimation of mse(a).}
Concrete evaluation of the $\mse(a)$ curves 
of (\ref{eq:mseofa}) 
shows
that the HL scheme typically is worthwhile, in that 
the mse is lower than that of the ML, for a range
of $a$ values. To find a good value of $a$ from 
data, a natural idea is to estimate the $\mse(a)$ 
and then pick its minimiser.  
For $\mse(a)$, the ingredients $\omega_\hl(a)$ 
and $\tau_{0,\hl}(a)$ involved in (\ref{eq:hatmuHLlimit})  
may be estimated consistently via plug-in 
of the relevant quantities. The difficulty
lies with the $\delta$ part, and more specifically 
with $\delta\delta^\tr$ in $\omega_\hl(a)\delta\delta^\tr\omega_\hl(a)$.
For this parameter, defined on the $O(1/\rootn)$ scale
via $\gamma=\gamma_0+\delta/\rootn$, the essential 
information lies in $D_n=\rootn(\hatt\gamma_\pl-\gamma_0)$, 
via parametric ML estimation in the extended $f(y,\theta,\gamma)$
model. As demonstrated and discussed 
in \citet[Chs.~6--7]{ClaeskensHjort08}, in connection
with construction of their Focused Information Criterion (FIC),
we have 
\beq
\label{eq:DntoD}
D_n\arr_d D\sim\N_r(\delta,Q), \quad 
   {\rm with\ }
  Q=J^{11}=(J_{11}-J_{10}J_{00}^{-1}J_{01})^{-1}. 
\eeq 
The factor $\delta/\rootn$
in the $O(1/\rootn)$ construction cannot be estimated consistently. 
Since $DD^\tr$ has mean $\delta\delta^\tr+Q$ in 
the 
limit,
we estimate squared bias 
parameters of the type $(b^\tr\delta)^2=b\delta\delta^\tr b$ using 
$\{b^\tr(D_nD_n^\tr-\hatt Q)b\}_+$, in which 
$\hatt Q$ estimates $Q=J^{11}$, and $x_+=\max(x,0)$. 
We construct the $r\times r$ matrix $\hatt Q$ 
from estimating and then inverting the full $(p+r)\times(p+r)$ 
Fisher information matrix $J_\wide$ of (\ref{eq:fisher}). 
This leads to estimating $\mse(a)$ using 
\beqn
\fic(a)=\{\hatt\omega_\hl(a)^\tr(D_nD_n^\tr-\hatt Q)\hatt\omega_\hl(a)\}_+
   +\hatt\tau_{0,\hl}(a)^2 
=\bigl[n\hatt\omega_\hl(a)^\tr\{(\hatt\gamma-\gamma_0)
   (\hatt\gamma-\gamma_0)^\tr
   -\hatt Q\}\hatt\omega_\hl(a)\bigr]_+
   +\hatt\tau_{0,\hl}(a)^2. 
\eeqn 

Figure \ref{figure:hel2}(b) displays such a root-fic curve, 
the estimated root-$\mse(a)$. 
Whereas the root-$\mse(a)$ curve shown in Figure \ref{figure:hel2}(a)
is coming from considerations and numerical investigation
of the extended $f(y,\theta,\gamma)$ model alone, 
pre-data, the root-$\fic(a)$ curve is constructed
for a given dataset. The start model and its extension
are as with Figure \ref{figure:hel2}(a), a $\Beta(\theta,1)$
inside a $\Beta(\theta,\gamma)$, with $n=100$ simulated
data points using $\gamma=1+\delta/\rootn$ with 
$\delta$ chosen as for Figure \ref{figure:hel2}(a). 
Again, the HL method was applied, using the second moment
$\psi=\E\,Y^2$ as both control and focus. 
The estimated risk is smallest for $a=0.41$. 


\saveustwolittle
\section{An illustration: Roman era Egyptian life-lengths}
\label{section:egypt}

A fascinating dataset on $n=141$ life-lengths from Roman era
Egypt, a century BC, is examined in \citet{Pearson1902},
where he compares life-length distributions 
from two societies, two thousand years apart. The data
are also discussed, modelled and analysed 
in \citet[Ch.~2]{ClaeskensHjort08}. 

\begin{figure}[h] 
\begin{center}
(a) \hspace*{6cm} (b) \\[-.3cm]
\includegraphics[width=7cm,height=6cm]{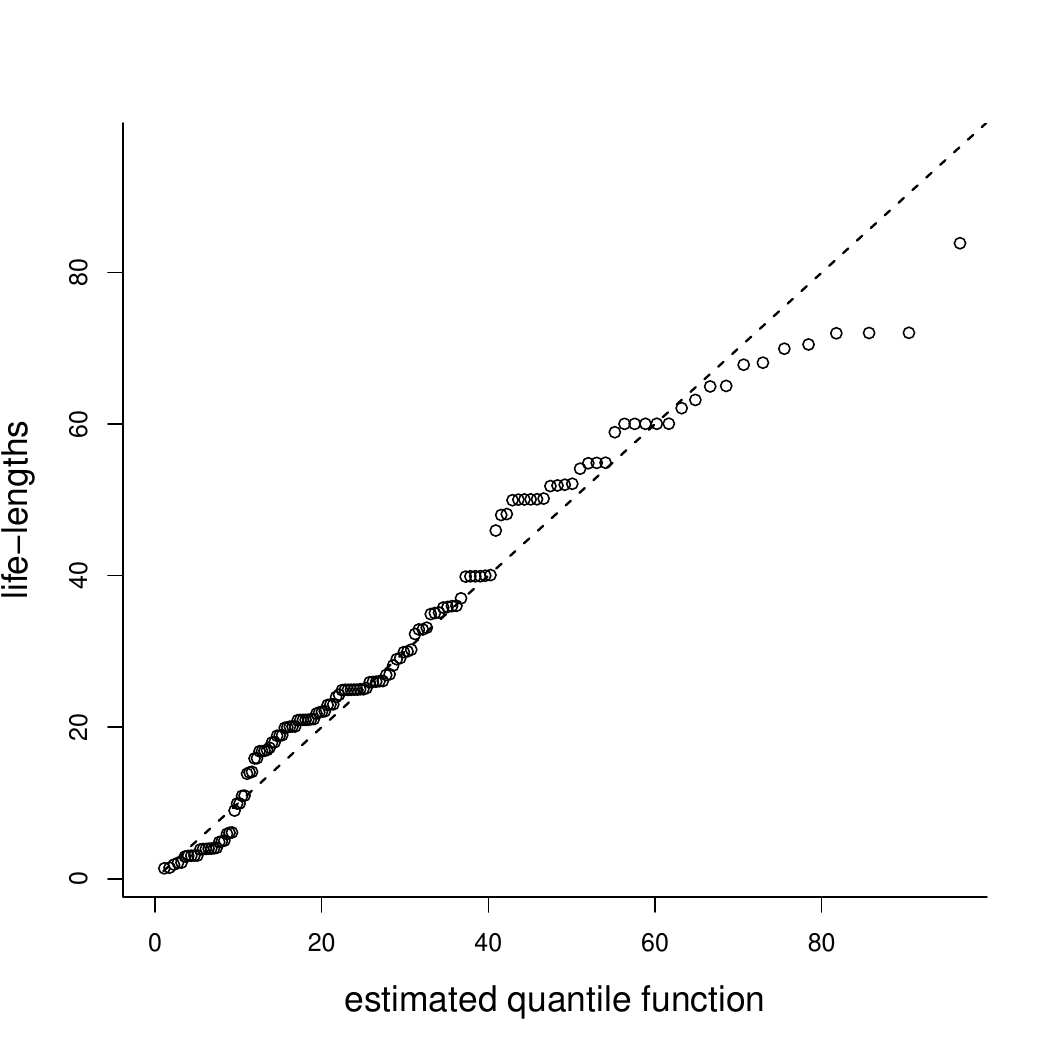} 
   \ \ \ \ \ \includegraphics[width=7cm,height=6cm]{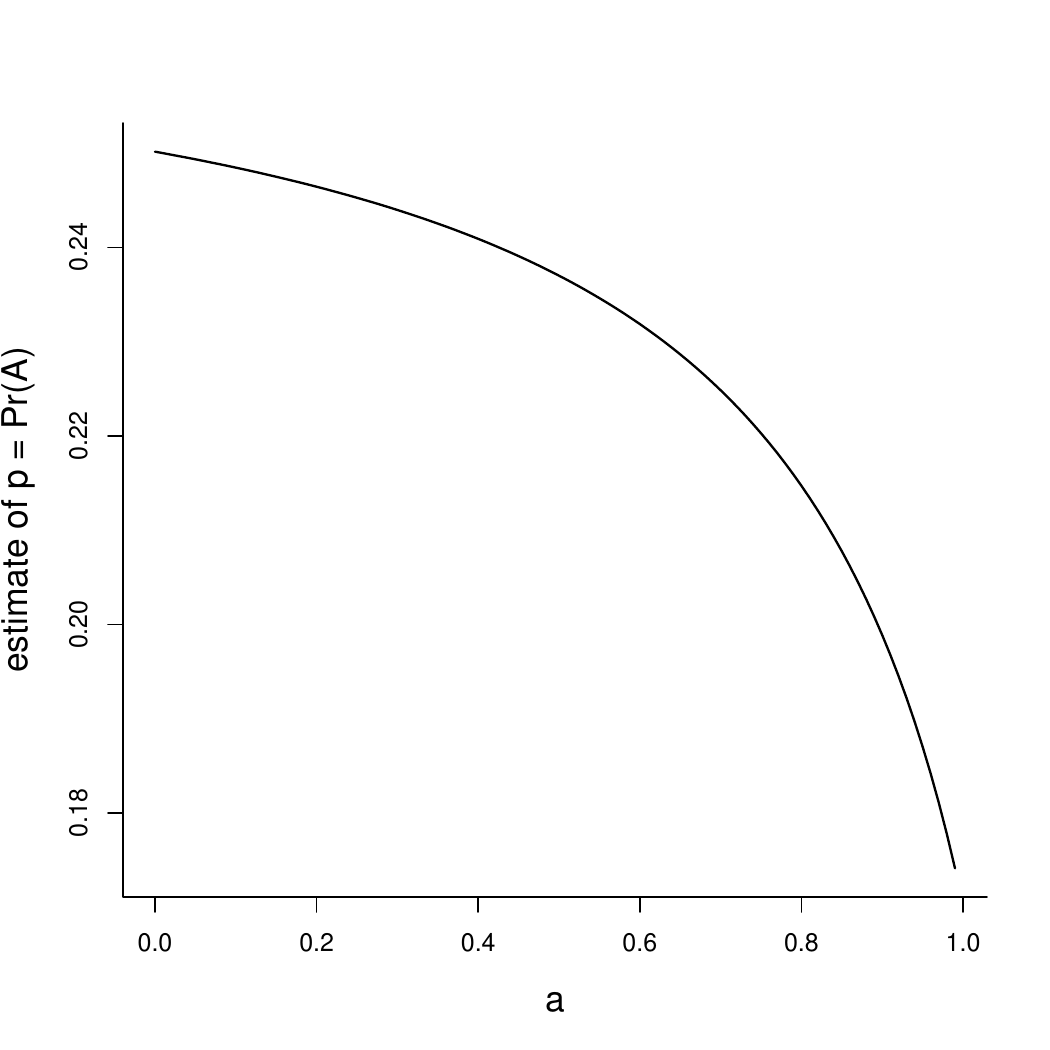}
\vspace*{-.3cm}
\end{center}
\caption{(a) The q-q plot shows the ordered life-lengths $y_{(i)}$
plotted against the ML-estimated gamma quantile function 
$F^{-1}(i/(n+1),\hatt b,\hatt c)$. 
(b) The curve $\hatt p_a$, with the probability 
$p=\Pr\{Y\in[9.5,20.5]\}$ estimated via the HL estimator, 
is displayed, as a function of the balance parameter $a$. 
At balance position $a=0.61$, the efficiency loss is 10\% compared to the ML 
precision under ideal gamma model conditions.}
\label{figure:hel5}
\end{figure}

\vspace{-0.75cm}
Here we have fitted the data to the $\Gamm(b,c)$ distribution,
first using the ML, with parameter estimates $(1.6077,0.0524)$.
The q-q plot of Figure \ref{figure:hel5}(a) displays the points 
$(F^{-1}(i/(n+1),\hatt b,\hatt c),y_{(i)})$, with 
$F^{-1}(\cdot,b,c)$ denoting the quantile function of the 
Gamma and $y_{(i)}$ the ordered life-lengths, from 1.5 to 96. 
We learn that the gamma distribution does a decent
job for these data, but that the fit is not good for the
longer lives. There is hence scope for the HL for estimating
and assessing relevant quantities in a more robust 
and indeed controlled fashion than via the ML. Here we focus on 
$p=p(b,c)=\Pr\{Y\in [L_1,L_2]\}=\int_{L_1}^{L_2}
   f(y,b,c)\,\dd y$, 
for age groups $[L_1,L_2]$ of interest. The hybrid log-likelihood 
is hence 
$h_n(b,c)=(1-a)\ell_n(b,c)+a\log R_n(p(b,c))$, 
with $R_n(p)$ being the EL associated
with $m(y,p)=I\{y\in[L_1,L_2]\}-p$. We may then, 
for each $a$, 
maximise this function and read off
both the HL estimates $(\hatt b_a,\hatt c_a)$ and the 
consequent $\hatt p_a=p(\hatt b_a,\hatt c_a)$. 
Figure \ref{figure:hel5}(b) displays this $\hatt p_a$,
as a function of $a$, 
for the age group $[9.5,20.5]$. For $a=0$ we have the 
ML based estimate $0.251$, and with increasing $a$
there is more weight to the EL,
which has the point estimate $0.171$. 

To decide on a good balance, recipes of 
Section \ref{section:finetuning} may be appealed to. 
The relatively speaking simplest of these is 
that associated with (\ref{eq:efficiencycontrol}),
where we numerically compute 
$\kappa_a=\{c^\tr(J^*)^{-1}K^*(J^*)^{-1}c\}^{1/2}$ 
for each $a$, at the ML position in the parameter space 
of $(b,c)$, and with $J^*$ and $K^*$ from (\ref{eq:JstarKstar}). 
The loss of efficiency $\kappa_a/\kappa_0$ is quite small
for small $a$, and is at level 1.10 for $a=0.61$. 
For this value of $a$, where confidence intervals are 
stretched 10\% compared to the gamma-model-based ML 
solution, we find $\hatt p_a$ equal to 0.232, 
with estimated standard deviation 
$\hatt\kappa_a/\rootn=0.188/\rootn=0.016$. 
Similarly the HL machinery may be put to work for 
other age intervals, for each such using the $p=\Pr\{Y\in[L_1,L_2]\}$
as both control and focus, and for models other than
the gamma. We may employ the HL with a collection 
of control parameters, like age groups, 
before landing on inference for a focus parameter; 
see Example \ref{example:example4}.
The more elaborate recipe of selecting $a$, 
developed in Section \ref{section:finetuning} and using $\fic(a)$,
can also be used here.

\vspace{-0.35cm}
\section{Further developments and the Supplementary Material} 
\label{section:concluding}

\saveusalittle 
Various concluding remarks and extra developments are placed 
in the article's Supplementary Material section. In particular,
proofs of Lemma \ref{lemma:lemma1}, Theorems 
\ref{theorem:theorem1} and \ref{theorem:theorem2} and Corollary \ref{cor:cor1} are
given there. Other material involves 
extension of the basic HL construction 
to regression type data, in Section \ref{subsection:regression};  
log-HL-profiling operations and a deviance fuction, 
leading to a full confidence curve for a focus parameter,
in Section \ref{subsection:cc}, 
an implicit goodness-of-fit test for the parametric vehicle model,
in Section \ref{subsection:goodnessoffit}, and  
a related but different hybrid likelihood construction,
in Section \ref{subsection:theothermethod}. 

\saveustwolittle 



\parindent0pt
\baselineskip12pt
\parskip4pt


\bibliographystyle{biometrika}
\bibliography{hmv_bibliography4}

\newpage

\setcounter{section}{19}
\renewcommand{\thesection}{\Alph{section}}

\parindent1.3pc 
\baselineskip20pt 
\parskip0pt
\setcounter{page}{1}

\section*{}
\label{section:SMpart}

\vspace*{-.5cm}

\centerline{\Large\bf Supplementary material to}
\centerline{\Large \bf `Hybrid combinations of parametric and empirical likelihoods'}

\centerline{Nils Lid Hjort, 
Ian W. McKeague, 
and Ingrid Van Keilegom} 
\centerline{\it University of Oslo, Columbia University, 
and KU Leuven} \vspace{.5cm}

This supplementary material contains the following sections.
Sections \ref{subsection:prooflemma1}, \ref{subsection:prooftheorem1},
\ref{subsection:proofcor1}, 
\ref{subsection:prooftheorem2} give the technical proofs
of Lemma \ref{lemma:lemma1}, Theorem \ref{theorem:theorem1},
Cororally \ref{cor:cor1} and Theorem \ref{theorem:theorem2}. 
Then Section \ref{subsection:regression} crucially indicates how 
the HL methodology can be lifted from the i.i.d.~case to regression
type models, whereas a Wilks type theorem based on HL-profiling, 
useful for constructing confidence curves for focus parameters,
is developed in Section \ref{subsection:cc}. 
An implicit goodness-of-fit test for the parametric working
model is identified in Section \ref{subsection:goodnessoffit}. 
Finally Section \ref{subsection:theothermethod} 
describes an alternative hybrid approach, related to, 
but different from the HL. This alternative method is first-order 
equivalent to the HL method inside $O(1/\rootn)$
neighbourhoods of the parametric vehicle model,
but not at farther distances.

\subsection{Proof of Lemma \ref{lemma:lemma1}}
\label{subsection:prooflemma1}

The proof is based on techniques and arguments related to those
of \citet{HMV09}, but with necessary extensions and modifications.

For the maximiser of $G_n(\cdot,s)$, write 
$\hatt\lambda_n(s)=\|\hatt\lambda_n(s)\| u(s)$ for a vector $u(s)$ of unit length.
With arguments as in \citet[p.~220]{Owen01}, 
$$\|\hatt\lambda_n(s)\| \big\{u(s)^\tr W_n(s) u(s)-E_n(s)u(s)^\tr V_n(s)\big\} \le u(s)^\tr V_n(s), $$
with $E_n(s)=n^{-1/2}\maxin\|m_{i,n}(s)\|$, which tends to zero
in probability uniformly in $s$ by assumption (iii). 
Also from assumption (i), $\sup_{s \in S} |u(s)^\tr V_n(s)|=O_\pr(1)$.  Moreover, 
$u(s)^\tr W_n(s) u(s) \ge e_{n,\min}(s)$, the smallest eigenvalue
of $W_n(s)$, which converges in probability to the smallest eigenvalue of $W$, 
and this is bounded away from zero by assumption (ii).
It follows that $\|\hatt\lambda_n(s)\|=O_\pr(1)$ uniformly in $s$. 
Also, $\lambda_n^*(s)=W_n(s)^{-1}V_n(s)$ is bounded in probability uniformly in $s$. 
Via $\log(1+x)=x-\half x^2+\third x^3h(x)$, 
where $|h(x)|\le2$ for $|x|\le\half$, write 
$$G_n(\lambda,s)=2\lambda^\tr V_n(s)-\half\lambda^\tr W_n(s)\lambda+r_n(\lambda,s)
   = G_n^*(\lambda,s)+r_n(\lambda,s). $$
For arbitrary $c>0$, consider any $\lambda$ with $\|\lambda\|\le c$. 
Then we find 
$$|r_n(\lambda,s)|\le{2\over 3}\sumin 
   |\lambda^\tr m_{i,n}(s)/\rootn|^3\,|h(\lambda^\tr m_{i,n}(s)/\rootn)| 
   \le {4\over 3}E_n(s)\|\lambda\|\lambda^\tr W_n(s)\lambda
   \le {4\over 3}E_n(s)c^3e_{n,\max}(s), $$
in terms of the largest eigenvalue of $W_n(s)$, 
as long as $cE_n(s)\le\half$. 
Choose $c$ big enough to have both $\hatt\lambda_n(s)$
and $\lambda_n^*(s)$ inside this ball for all $s$ with probability
exceeding $1-\eps'$, for a preassigned small $\eps'$. Then,
\beqn
&& \Pr\Big(\sup_{s \in S} |\max_\lambda G_n(\lambda,s)
   -\max_\lambda G_n^*(\lambda,s)|\ge\eps\Big) \\
&& \le \Pr\Big(\sup_{s \in S} \sup_{\|\lambda\| \le c} |G_n(\lambda,s) 
   - G_n^*(\lambda,s)| \ge \eps\Big) \\
&& \le\Pr\Big((4/3) c^3 \sup_{s \in S} (E_n(s) e_{n,\max}(s)) \ge\eps\Big) 
+ \Pr\Big(\sup_{s \in S}\|\hatt\lambda_n(s)\|>c\Big) \\
&&\qquad\qquad 
+ \Pr\Big(\sup_{s \in S}\|\lambda_n^*(s)\|>c\Big) + \Pr\Big(c\sup_{s \in S} E_n(s) 
   > \half\Big). 
\eeqn 
The lim-sup of the probability sequence on the left 
hand side is hence bounded by $4\eps'$. We have proven that 
$\sup_{s \in S} |\max_\lambda G_n(\lambda,s)
   -\max_\lambda G_n^*(\lambda,s)| \arr_\pr 0$. \qed

\subsection{Proof of Theorem \ref{theorem:theorem1}}
\label{subsection:prooftheorem1}

We work with the two components of (\ref{eq:Anfunction}) 
separately. First, with $U_n=n^{-1/2}\sumin u(Y_i,\theta_0)$,
which tends to $U_0\sim\N_p(0,J)$, cf.~(\ref{eq:CLTatwork}), 
\beq
\label{eq:loglikpart}
\ell_n(\theta_0+s/\rootn)-\ell_n(\theta_0)
   =s^\tr U_n-\half s^\tr Js+\eps_n(s), 
   \quad {\rm with\ }\sup_{s\in S}|\eps_n(s)|\arr_\pr 0, 
\eeq 
under various sets of mild regularity conditions. 
If $\log f(y,\theta)$ is concave in $\theta$, no 
other conditions are required, beyond finiteness of 
the Fisher information matrix $J$, see \citet{HjortPollard94}. 
Without concavity, but assuming the existence of third order derivatives 
$D_{i,j,k}(y,\theta)=\dell^3\log f(y,\theta)/\dell\theta_i\dell\theta_j\dell\theta_k$,
it is straightforward via Tayor expansion to verify (\ref{eq:loglikpart}) 
under the condition that $\sup_{\theta\in{\cal N}}\max_{i,j,k}|D_{i,j,k}(Y,\theta)|$ 
has finite mean, with $\cal N$ a neighbourhood around $\theta_0$. 
This condition is met for most of the usually employed 
parametric families. We finally point out that (\ref{eq:loglikpart})
can be established without third order derivatives, 
with a mild continuity condition on the second derivatives,
see e.g.~\citet[Ch.~18]{Ferguson96}.  

Secondly, we shall see that Lemma \ref{lemma:lemma1} may be applied, 
implying 
\beq
\label{eq:quadraticfriend}
\log R_n(\mu(\theta_0+s/\rootn))
   =-\half V_n(s)^\tr W_n(s)^{-1}
   V_n(s)+o_\pr(1),
\eeq 
uniformly in $s \in S$. 
For this to be valid it is in view of Lemma \ref{lemma:lemma1}
sufficient to check  condition (i) of that lemma (we assumed conditions (ii) and (iii)).
Here (i) follows using (\ref{eq:msmooth1}), since
\beqn
\sup_s \|V_n(s)\|   = \sup_s \|V_n(0)+\xi_ns\| +o_\pr(1)  \arr_d \sup_s \|V_0+\xi_0 s\|.  
\eeqn
Hence, $\sup_s \|V_n(s)\| = O_\pr(1)$.  

From these efforts we find 
\beqn
\log R_n(\mu(\theta_0+s/\rootn))-\log R_n(\mu(\theta_0))
   &\arr_d& -\half (V_0+\xi_0 s)^\tr W^{-1}(V_0+\xi_0s)
   +\half V_0^\tr W^{-1} V_0 \\
   &=&-V_0^\tr W^{-1}\xi_0s-\half s^\tr\xi_0^\tr W^{-1}\xi_0s. 
\eeqn 
This convergence also takes place jointly with 
(\ref{eq:loglikpart}), in view of (\ref{eq:CLTatwork}), 
and we arrive at the conclusion of the theorem. \qed

\subsection{Proof of Corollary \ref{cor:cor1}}
\label{subsection:proofcor1}

Corollary \ref{cor:cor1} is valid under the following conditions, where 
$\Gamma(\cdot)$ is defined in (\ref{eq:Gammafunction}): 
\begin{itemize}
\item[(A1)] For all $\eps>0$, 
$\sup_{\|\theta-\theta_0\| > \eps} \Gamma(\theta) < \Gamma(\theta_0)$.
\item[(A2)] The class $\big\{y\mapsto\frac{\dell}{\dell\theta} \log f(y,\theta)
   \colon\theta \in \Theta \big\}$ is $P$-Donsker 
   (see e.g.~\citet[Ch.~2]{vdVaartWellner96}).
\item[(A3)] Conditions (C0)--(C2) and (C4)--(C6) in 
\citet{Lopezetal09} are valid, with their function 
$g(X,\mu_0,\nu)$ replaced by our function $m(Y,\mu(\theta))$,
with $\theta$ playing the role of $\nu$, 
except that instead of demanding boundedness of our 
function $m(Y,\mu)$ we assume merely that the class 
$$ y \mapsto \frac{m(y,\mu) m(y,\mu)^\tr}{\{1 + \xi^\tr m(y,\mu)\}^2},$$ 
with $\mu$ and $\xi$ in a neighbourhood of $\mu(\theta_0)$ 
and $0$, is $P$-Donsker (this is a much milder condition
than boundedness). 
\end{itemize}

First note that $\Gamma_n(\theta)$ can be written as 
$$ \Gamma_n(\theta) = (1-a) \, 
  n^{-1} \sumin\{\log f(Y_i,\theta) - \log f(Y_i,\theta_0)\}
 - a\, n^{-1} \sumin \log\big(1+\hatt\xi(\theta)^\tr m(Y_i,\mu(\theta))\big), $$
where $\hatt\xi(\theta)$ is the solution of 
$$ n^{-1} \sumin \frac{m(Y_i,\mu(\theta))}{1+\xi^\tr m(Y_i,\mu(\theta))} = 0. $$
Note that this corresponds with the formula of $\log R_n$ 
given below Lemma \ref{lemma:lemma1} but with 
$\lambda(\theta)/\rootn$ relabelled as $\xi(\theta)$.
That the $\hatt\xi(\theta)$ solution is unique follows 
from considerations along the lines of \citet[p.~415]{Lopezetal09}. 
To prove the consistency part, 
we make use of Theorem 5.7 in \citet{vdV}.  
It suffices by condition (A1) to show that 
$\sup_\theta |\Gamma_n(\theta)- \Gamma(\theta)| \arr_\pr 0$, 
which we show separately for the ML and the EL part. 
For the parametric part we know that 
$n^{-1} \ell_n(\theta) - \E \log f(Y,\theta)$ is $o_\pr(1)$ 
uniformly in $\theta$ by condition (A2). For the EL part, 
the proof is similar to the proof of Lemma 4
in \citet{Lopezetal09} (except that no rate 
is required here and that the convergence
is uniformly in $\theta$), and hence details are omitted. 
 
Next, to prove statement (ii) of the corollary, 
we make use of Theorems 1 and 2 in \citet{Sherman93} 
about the asymptotics for the maximiser of 
a (not necessarily concave) criterion function, 
and the results in \citet{Lopezetal09}, who 
use the \citet{Sherman93} paper to establish asymptotic normality 
and a version of the Wilks theorem in an EL context with nuisance parameters. 
For the verification of the conditions of Theorem 1 
(which shows root-$n$ consistency of $\hatt\theta_\hl$) 
and Theorem 2 (which shows asymptotic normality of $\hatt\theta_\hl$) 
in \citet{Sherman93}, we consider separately the ML part 
and the EL part. We note that Theorem 1 in \citet{Sherman93} 
requires consistency of the estimator, which we here
have established by arguments above. 
For the EL part all the work is already done using our Theorem \ref{theorem:theorem1}
and Lemmas 1--6 in \citet{Lopezetal09}, which are 
valid under condition (A3). Next, the conditions of Theorems 1 and 2 in 
\citet{Sherman93} for the ML part follow using standard arguments from 
parametric likelihood theory and condition (A2).  
It now follows that $\hatt\theta_\hl$ is asymptotically normal, 
and its asymptotic variance is equal to $(J^*)^{-1} K^* (J^*)^{-1}$ using 
Theorem \ref{theorem:theorem1}. 

Finally, claim (iii) of the corollary follows from a combination of 
Theorem \ref{theorem:theorem1} with $s=\rootn (\hatt\theta_\hl-\theta_0)$ 
and the asymptotic normality of $\rootn (\hatt\theta_\hl-\theta_0)$ 
to $(J^*)^{-1} U^*$. Indeed,
$$ 2\{h_n(\hatt\theta_\hl)-h_n(\theta_0)\} \arr_d 
   2 \{(U^*)^\tr (J^*)^{-1} U^* 
   - \half (U^*)^\tr (J^*)^{-1} J^* (J^*)^{-1} U^* \}  = (U^*)^\tr (J^*)^{-1} U^*, $$
and this finishes the proof of the corollary. \qed


\subsection{Proof of Theorem \ref{theorem:theorem2}}
\label{subsection:prooftheorem2}

To prove Theorem \ref{theorem:theorem2}, 
we revisit several previous arguments for the 
$A_n(\cdot)\arr_d A(\cdot)$ part of Theorem~\ref{theorem:theorem1}, 
but now needing to extend these to the case of the model 
departure parameter $\delta$ being present. First, we have 
$$\ell_n(\theta_0+s/\rootn)-\ell_n(\theta_0)
   =U_ns-\half s^\tr J_ns+o_\pr(1)
   \arr_d (U+J_{01}\delta)^\tr s-\half s^\tr J_{00}s. $$
This is essentially since 
$U_n=n^{-1/2}\sumin u(Y_i,\theta_0)$ now is seen to 
have mean $J_{01}\delta$, but the same variance, up 
to the required order. We need a parallel result for 
$V_{n,0}=n^{-1/2}\sumin m(Y_i,\mu(\theta_0))$ under $f_\true$.
Here 
\beqn
\E_\true\,m(Y,\mu(\theta_0))
&=&\int m(y,\mu(\theta_0))
   f(y,\theta_0)\{1+S(y)^\tr\delta/\rootn+o(1/\rootn)\}\,\dd y \\
&=&0+K_{01}\delta/\rootn+o(1/\rootn),
\eeqn
yielding $V_{n,0}\arr_d V_0+K_{01}\delta$. 
Along with some further details, this leads to 
the required extension of the $A_n\arr_d A$ part of 
Theorem \ref{theorem:theorem1} and its proof, 
to the present local neighbourhood model state of affairs; 
$$A_n(s)=h_n(\theta_0+s/\rootn)-h_n(\theta_0)
   \arr_d A(s)=s^\tr U^*_\plus-\half s^\tr J^*s, $$
with $J^*$ as defined earlier and with 
$$U^*_\plus=(1-a)(U+J_{01}\delta)-a\xi_0^\tr W^{-1}(V_0+K_{01}\delta)
   =U^*+L_{01}\delta. $$
Following and then modifying the technical details
of the proof of Corollary \ref{cor:cor1}, we arrive at 
\beqn
\rootn(\hatt\theta_\hl-\theta_0)
   \arr_d(J^*)^{-1}(U^*+L_{01}\delta) 
   \sim\N_p((J^*)^{-1}L_{01}\delta,(J^*)^{-1}K^*(J^*)^{-1}),
\eeqn
as required. \qed

\subsection{The HL for regression models}
\label{subsection:regression}

Our HL machinery can be lifted from the iid framework
to regression. The following example illustrates
the general idea. Consider the normal linear regression
model for data $(x_i,y_i)$, with covariate vector
$x_i$ of dimension say $d$, and with $y_i$ having mean
$x_i^\tr\beta$. The ML solution is associated with
the estimation equation $\E\,m(Y,X,\beta)=0$, where
$m(y,x,\beta)=(y - x^\tr\beta) x$. The underlying regression
parameter can be expressed as
$\beta=(\E\,XX^\tr)^{-1}\,\E\,X Y$, involving also
the covariate distribution. Consider now a subvector
$x_0$, of dimension say $d_0<d$, and the associated
estimating equation $m_0(y,x,\gamma)=(y - x_0^\tr\gamma) x_0$.
This invites the HL construction
$(1-a)\ell_n(\beta) + a\log R_n(\gamma(\beta))$.
Here $\ell_n(\beta)$ is the ordinary parametric log-likelihood;
$R_n(\gamma)$ is the EL associated with $m_0$;
and $\gamma(\beta)$ is $(\E\,X_0X_0^\tr)^{-1}\,\E\,X_0 Y$
seen through the lens of the smaller regression, where
$\E\,X_0 Y=X_0 X^\tr\beta$. This leads to inference
about $\beta$ where it is taken into account
that regression with respect to the $x_0$ components
is of particular importance. 

\subsection{Confidence curve for a focus parameter}
\label{subsection:cc}

For a focus parameter $\psi=\psi(\theta)$, consider the profiled 
log-hybrid-likelihood function 
$h_{n,\prof}(\psi)=\max\{h_n(\theta)\colon\allowbreak\psi(\theta)=\psi\}$. 
Note that $h_{n,\max}=h_n(\hatt\theta_\hl)$ is also the same 
as $h_{n,\prof}(\hatt\psi_\hl)$. We shall find use for 
the hybrid deviance function associated with $\psi$, 
$$\Delta_n(\psi)=2\{h_{n,\prof}(\hatt\psi_\hl)-h_{n,\prof}(\psi)\}. $$ 
Essentially relying on Theorem \ref{theorem:theorem1}, 
which involves matrices $J^*$ and $K^*$ and the limit
variable $U^*\sim\N_p(0,K^*)$, we show below that 
\beq
\label{eq:Deltaconvergence}
\Delta_n(\psi_0)\arr_d 
   \Delta={\{c^\tr (J^*)^{-1}U^*\}^2\over c^\tr (J^*)^{-1}c}
   \sim k\chi^2_1, 
\eeq 
where $k=c^\tr(J^*)^{-1}K^*(J^*)^{-1}c/c^\tr(J^*)^{-1}c$.
Here $c=\dell\psi(\theta_0)/\dell\theta$, as in (\ref{cor:cor3}). 
Estimating this $k$ via plug-in then leads to the full confidence curve 
$\cc(\psi)=\Gamma_1(\Delta_n(\psi)/\hatt k)$, see 
\citet[Chs.~2--3]{CLP16}, often improving on the usual
symmetric normal-approximation based confidence intervals. 
Here $\Gamma_1(\cdot)$ is the distribution function of the $\chi^2_1$. 

To show (\ref{eq:Deltaconvergence}), we go via 
a profiled version of $A_n(s)$ in (\ref{eq:Anfunction}), namely 
$B_n(t)=h_{n,\prof}(\psi_0+t/\rootn)-h_{n,\prof}(\psi_0)$, 
where $\psi_0=\psi(\theta_0)$. For $B_n(t)$ and $\Delta_n(\psi)$
we have the following. 

\begin{theorem}
\label{theorem:theorem3}
Assume the conditions of Theorem \ref{theorem:theorem1} 
are in force. With $\psi_0=\psi(\theta_0)$ the true parameter value, 
and $c=\dell\psi(\theta_0)/\dell\theta$, we have 
$B_n(t)\arr_d B(t)=\{c^\tr (J^*)^{-1}U^*t-\half t^2\}/c^\tr (J^*)^{-1}c$. 
Also, 
$$\Delta_n(\psi_0)=2\,\max B_n 
   \arr_d \Delta=2\,\max B={\{c^\tr (J^*)^{-1}U^*\}^2\over c^\tr (J^*)^{-1}c}. $$
\end{theorem}

It is clear that $\Delta\sim k\chi^2_1$, with the $k$ given above. 
Proving the theorem is achieved via Theorem \ref{theorem:theorem1},
along the lines of a similar type of result for log-likelihood
profiling given in \citet[Section 2.4]{CLP16}, and we 
leave out the details. 

\begin{remark}
\label{remark:azero}
{{\rm 
The special case of $a=0$ for the HL construction 
corresponds to parametric ML estimation,
and results reached above specialise to the classical results 
$\rootn(\hatt\theta_\pl-\theta_0)\arr_d\N_p(0,J^{-1})$, 
$2\{\ell_{n,\max}-\ell_n(\theta_0)\}\arr_d\chi^2_p$, and 
$\rootn(\hatt\psi_\pl-\psi_0)\arr_d\N(0,c^\tr J^{-1}c)$. 
Theorem \ref{theorem:theorem3} is then the Wilks theorem
for the profiled log-likelihood function. 
The other extreme case is that of $a\arr1$, 
with the EL applied to $\mu=\mu(\theta)$. 
Here Theorem \ref{theorem:theorem1} yields 
$U^*=-\xi_0^\tr W^{-1}V_0$, and with both $J^*$ and $K^*$
equal to $\xi_0^\tr W^{-1}\xi_0$. This case corresponds
to a version of the classic EL chi-squared 
result, now filtered through the parametric model, and with 
$-2\log R_n(\mu(\theta_0))\arr_d (U^*)^\tr (J^*)^{-1}U^*\sim\chi^2_p$.
Also, $\rootn(\hatt\psi_\el-\psi_0)\arr_d\N(0,\kappa^2)$,
with $\kappa^2=c^\tr\xi_0^\tr W^{-1}\xi_0c$;
here $\hatt\psi_\el=\psi(\hatt\theta_\el)$ in terms of the 
EL estimator, the maximiser 
of $R_n(\mu(\theta))$. 
}} 
\end{remark}

\subsection{An implied goodness-of-fit test for the parametric model}
\label{subsection:goodnessoffit} 

Methods developed in Section \ref{section:finetuning},
in particular those associated with estimating the 
mean squared error of the final estimator, lend
themselves nicely to a goodness-of-fit test for 
the parametric working model, as follows. 
We accept the parametric model if the $\fic(a)$
criterion of Section \ref{section:finetuning} 
tells us that $\hatt a=0$ is the best
balance, and if $\hatt a>0$ the model is rejected. 
This model test can be accurately examined, by 
working out an expression for the derivative of 
$\fic(a)$ at $a=0$, say $\hatt Z_n^0$; we reject the model 
if $\hatt Z_n^0>0$ (since then and only then is $\hatt a$ positive). 

Here $\hatt Z_n^0$ is the estimated version of 
the limit experiment variable $Z^0$, which we shall
identify below, as a function of $D\sim\N_q(\delta,Q)$,
cf.~(\ref{eq:DntoD}). Let us write 
$\omega_{\hl}(a)=\omega+a\nu+O(a^2)$. Since 
$\tau_{0,\hl}(a)^2=\tau_0^2+O(a^2)$, the derivative of 
$$\fic(a)=(\omega+a\nu)^\tr (DD^\tr-Q)(\omega+a\nu)
   +\tau_0^2+O(a^2) $$
with respect to $a$, at zero, is seen to be 
$Z^0=2\omega^\tr(DD^\tr-Q)\nu$. Hence the limit experiment
version of the test is to reject the parametric model
if $(\omega^\tr D)(\nu^\tr D)>\omega^\tr Q\nu$, or 
$$Z={\omega^\tr D\over (\omega^\tr Q\omega)^{1/2}}
    {\nu^\tr D\over (\nu^\tr Q\nu)^{1/2}}
   >\rho={\omega^\tr Q\nu\over
        (\omega^\tr Q\omega)^{1/2}(\nu^\tr Q\nu)^{1/2}}. $$  
Under the null hypothesis of the model, $Z$
is equal in distribution to $X_1X_2$, where $(X_1,X_2)$
is a binormal pair, with zero means, unit variances,
and correlation $\rho$. The implied significance level,
of the implied goodness of fit test, is hence 
$\alpha=\Pr\{X_1X_2>\rho\}$, which can be read off via
numerical integration or simulation, for a given $\rho$. 

The $\nu$ quantity can be identified with a bit of algebraic
work, and then estimated consistently from the data. 
We note that for the special case of $m(y,\mu)=g(y)-\mu$, 
and with focus on this mean parameter $\mu=\E\,g(Y)$, 
then $\nu$ becomes proportional to $\omega$. The 
test above is then equivalent to rejecting the model
if $(\hatt\omega^\tr D_n)^2/\hatt\omega^\tr\hatt Q\hatt\omega>1$,
which under the null model happens with probability 
converging to $\alpha=\Pr\{\chi^2_1>1\}=0.317$.  


\subsection{A related hybrid estimation method}
\label{subsection:theothermethod}

In earlier sections we have motivated and developed theory 
for the hybrid likelihood and the HL estimator. 
A crucial factor has been the quadratic approximation 
(\ref{eq:quadraticfriend}). The latter is essentially
valid within a $O(1/\rootn)$ neighbourhood around the 
true data generating mechanism, and has yielded 
the results of Sections \ref{section:basictheory} 
and \ref{section:finetuning}. 

A related though different strategy is however to 
take this quadratic approximation as the starting point.
The suggestion is then to define the alternative hybrid estimator
as the maximiser $\tilda\theta$ of 
\beq
\label{eq:newmethod}
N_n(\theta)=(1-a)\ell_n(\theta)
   -\half a V_n(\theta)^\tr W_n(\theta)^{-1}V_n(\theta). 
\eeq 
Under and close to the parametric working model, 
the HL estimator $\hatt\theta$ and the 
new-HL estimator $\tilda\theta$ are first-order 
equivalent, in the sense of $\rootn(\hatt\theta-\tilda\theta)\arr_\pr0$. 
Of course we could have put up (\ref{eq:newmethod}) 
without knowing or caring about EL or HL in the first 
place, and with different balance weights. 
But here we are naturally led to the balance weights $1-a$ 
for the log-likelihood and $-\half a$ for the quadratic 
form, from the HL construction.

The advantage of (\ref{eq:newmethod}) is partly that 
it is easier computationally, without a layer of 
Lagrange maximisation for each $\theta$. More importantly,
it manages well also outside the $O(1/\rootn)$ 
neighbourhoods of the working model. The new-HL 
estimator tends under weak regularity conditions to 
the maximiser $\theta_0$ of the limit function of 
$N_n(\theta)/n$, which may written 
$$N(\theta)=(1-a)\int g\log f_\theta\,\dd y
   -\half a\,v_\theta^\tr(\Sigma_\theta+v_\theta v_\theta^\tr)^{-1}v_\theta, $$
in terms of $v_\theta=\E_f\,m(Y,\mu(\theta))$ 
and $\Sigma_\theta=\Var_f m(Y,\mu(\theta))$. Note next that 
$(A+xx^\tr)^{-1}=A^{-1}-A^{-1}xx^\tr A^{-1}/(1+x^\tr A^{-1}x)$,
for invertible $A$ and vector $x$ of appropriate dimension.
This leads to the identity 
$$x^\tr(A+xx^\tr)^{-1}x={x^\tr A^{-1}x\over 1+x^\tr A^{-1}x}. $$
Hence the $\theta_0$ associated with the new-HL method
is the miminiser of the statistical distance function 
\beq
\label{eq:divergence}
d_a(f,f_\theta)=(1-a)\KL(f,f_\theta)
   +\half a{v_\theta^\tr \Sigma_\theta^{-1} v_\theta
   \over 1+v_\theta^\tr \Sigma_\theta^{-1} v_\theta} 
\eeq 
from the real $f$ to the modelled $f_\theta$; 
here $\KL(f,f_\theta)=\int f\log(f/f_\theta)\,\dd y$ is 
the Kullback--Leibler distance. For $a$ close to zero,
the new-HL is essentially maximising the log-likelihood 
function, associated with attempting to minimise
the KL divergence. For $a$ coming close to 1 
the method amounts to minimising an empirical version
of $v_\theta^\tr\Sigma_\theta^{-1}v_\theta$, which means
making $v_\theta=\E_f\,m(Y,\mu(\theta))$ close to zero.
This is also what the empirical likelihood is aiming at. 


\vspace*{.5cm}

\noindent
{\bf \large References}

\bib Ferguson, T.S. (1996). {\em A Course in Large Sample Theory}. Chapman \& Hall, Melbourne.

\bib Hjort, N.L., McKeague, I.W. and Van Keilegom, I. (2009). Extending the scope of empirical likelihood. {\it Annals of Statistics}, {\bf 37}, 1079--1111.

\bib Hjort, N.L. and Pollard, D. (1994). Asymptotics for minimisers of convex processes. Technical report, Department of Mathematics, University of Oslo.

\bib Molanes L\'opez, E., Van Keilegom, I. and Veraverbeke, N. (2009). Empirical likelihood for non-smooth criterion function. {\it Scandinavian Journal of Statistics}, {\bf 36}, 413--432.

\bib Owen, A. (2001). {\em Empirical Likelihood}. Chapman \& Hall/CRC, London.

\bib Schweder, T. and Hjort, N.L. (2016). {\em Confidence, Likelihood, Probability: Statistical Inference with Confidence Distributions}. Cambridge University Press, Cambridge.

\bib Sherman, R.P. (1993). The limiting distribution of the maximum rank correlation estimator. {\it Econometrica}, {\bf 61}, 123--137.

\bib van der Vaart, A.W. (1998). {\em Asymptotic Statistics}. Cambridge University Press, Cambridge.

\bib van der Vaart, A.W. and Wellner, J.A. (1996). {\em Weak Convergence and Empirical Processes}. Springer-Verlag, New York.

\end{document}